\documentclass[11pt]{article}

\usepackage{natbib}
\usepackage{amsmath, amsthm, amssymb, bbm, setspace,bigints}
 \usepackage[margin=1 in]{geometry}
\usepackage{xcolor}
\usepackage{caption}
\usepackage{multirow}
\usepackage{adjustbox}
\usepackage{textcomp}
\usepackage{graphicx}
\usepackage{subfigure}
\usepackage{float}
\usepackage{algorithm}  
\usepackage{algpseudocode}
\usepackage{bm}
\usepackage{natbib}
\setcitestyle{authoryear,round}

\bibpunct[, ]{[}{]}{,}{n}{,}{,}
\renewcommand\bibfont{\fontsize{10}{12}\selectfont}

\theoremstyle{plain}

\newtheorem{proposition}{Proposition}

\theoremstyle{definition}
\newtheorem{definition}{Definition}

\theoremstyle{remark}
\newtheorem{remark}{Remark}

\begin{document}


\title{Depth-Based Statistical Inferences in the Spike Train Space}

\author{Xinyu Zhou, Wei Wu \\ Department of Statistics, Florida State University}

	\maketitle

\begin{abstract}
Metric-based summary statistics such as mean and covariance have been introduced in neural spike train space. They can properly describe template and variability in spike train data, but are often sensitive to outliers and expensive to compute. Recent studies also examine outlier detection and classification methods on point processes. These tools provide reasonable and efficient result, whereas the accuracy remains at a low level in certain cases. In this study, we propose to adopt a well-established notion of statistical depth to the spike train space.  This framework can naturally define the median in a set of spike trains, which provides a robust description of the `center' or `template' of the observations. It also provides a principled method to identify `outliers' in the data and classify data from different categories.  We systematically compare the median with the state-of-the-art `mean spike trains' in terms of robustness and efficiency. The performance of our novel outlier detection and classification tools will be compared with previous methods. The result shows the median has superior description for `template' than the mean. Moreover, the proposed outlier detection and classification perform more accurately than previous methods. The advantages and superiority are well illustrated with simulations and real data.
\end{abstract}

\section{Introduction}

Modeling and analysis of neuronal spike trains has been a central topic in computational neuroscience.  A commonly used model is the point process framework, where each spike train is considered as a realization of a temporal point process (i.e., a list of event times in increasing order). 
\citet{kass2001spike} first introduced homogeneous Poisson process model to construct time dependent probability model for a sample of spike trains; \citet{brown2003likelihood} adopted maximum likelihood method to study the firing rate of spike train samples.  
\citet{truccolo2005point} developed a generalized linear model framework where spike history and external covariate are well incorporated to characterize firing activity at each time point. 
All these point process models and their applications in modeling spiking activity in various brain areas were thoroughly summarized in  
\citet{kass2005statistical,kass2018computational}. Although the spike trains have been studied for decades, modeling and inferences in the space of spike trains are still very limited.  For example, very few studies have addressed the summary statistics of the sample of spike trains. Common questions such as `what is the template in a given set of observations and how much is the variability in the sample?' are still not well examined in the field. In this case, each single spike train is treated as an object in the spike train space and the classical summary statistics, such as mean and variance, can be directly defined to the given spike train sample. 



Since the space of spike trains is not a conventional vector space, the Euclidean metrics cannot be directly applied. Two basic constraints are: 1) the cardinality (i.e., number of spikes) of each trial varies within a sample, and 2) the spike times are in increasing order.  Non-Euclidean metrics have been developed to measure dissimilarities between spike trains.  In a pioneer investigation, 
\citet{victor1997metric} introduced a proper metric on spike trains, which corresponds to a Manhattan distance in the $\mathbb L^1$ space.  \citet{wu2011information} generalized the metric definition to the $\mathbb L^p$ space $(p \ge 1)$. When $p=2$, a Euclidean metric was introduced and the notion of mean spike train was naturally defined with the conventional minimum sum of squares.  Computational algorithms were then introduced to estimate the sample mean spike train \citep{wu2013estimating}.  However, we point out that 
 a common issue of the mean spike train is its non-robustness with respect to outliers in the given data.  This may significantly reduce the effectiveness of the mean spike train in practical use. 

In this paper, we will propose a robust measure of centrality, median spike train, to a sample of spike trains via the notion of statistical depth. Median is a centrality measure for any objects under a given ranking method. For spike trains in non-Euclidean space, we adopt the notion of statistical depth in the spike train space. Depth has been a powerful tool to measure the center-outward rank of multivariate and functional data in multiple decades. The notion of depth was first introduced by \citet{tukey1975mathematics} for multivariate data and the corresponding mathematical properties was thoroughly examined by \citet{zuo2000general}. Since the introduction, depth has been developed in many different forms, which include simplicial depth \citep{liu1988notion}, Mahalanobis depth \citep{liu1993quality}, zonoid depth \citep{dyckerhoff1996zonoid}, and projection depth \citep{zuo2000note}. Recently study of depth has focused on functional data. \citet{lopez2009concept} first introduced the concept of functional depth. The mathematical properties of functional depth were extensively studied by \citet{nieto2011properties} and the corresponding applications were about functional outlier detection  \citep{sun2011functional} and functional classification \citep{makinde2019classification}. 

The latest studies on depth have started to focus on point processes.  \citet{liu2017generalized} first introduced the notion of depth to spike train by using a generalized Mahalanobis depth, whereas the method ignores the orderingness of the spike events over time.  \citet{qi2021dirichlet} properly addressed the orderingness by the equivalent inter-spike interval representation in a simplex.  However, the simplex domain lacks important symmetric property and the proposed Dirichlet depth may not build a proper center-outward rank in practical use. To overcome these problems, \citet{zhou2023statistical} constructed a principled  Isometric Log-Ratio (ILR) depth for point process data.  By adopting the ILR transformation on the inter-spike intervals, each spike train is bijectively mapped to a vector in a multivariate Euclidean space.  This ILR depth is based on the Gaussian-like density function in this Euclidean space, which properly introduces symmetry and provides a center-outward rank as in the classical Mahalanobis depth. 
In this paper, we will adopt the ILR depth to construct the framework of spike train ranking. 

Based on the ranking by the depth value, the one with the largest depth is naturally the median spike train.  In addition, we can identify spike trains with low depth values as outliers when a proper threshold is well defined.  This can be done for a spike train with any cardinality.  We point out that  
outlier detection in spike trains is a relatively under-explored area as most outlier studies have focused on multivariate or functional data. \citet{liu2021event} developed one abnormal-event detection method based on Bayesian decision theory and hypothesis testing. This method can only identify unexpected occurrence or absence of spike events within a given trial, but not be used to evaluate whether a spike train itself is an outlier.  To detect the anomaly spike train, \citet{ojeda2019patterns} designed a clustering algorithm for Poisson process observations and \citet{zhu2020adversarial} combined Long Short-Term Memory network with marked spatio-temporal point process model to detect weak signal. However, these methods are limited within certain point process models and cannot be extended to more general cases. In a recent study, \citet{shchur2021detecting} adopted Out-of-Distribution detection with a newly proposed sum-of-squared-spacings (3S) statistic to detect anomalous point processes. This method is theoretically applicable to any point process data. In this paper, we will systematically compare our proposed outlier detection method with this 3S method. 


In addition to proposing the definition of median spike train and a new procedure for outlier detection, we will also develop a depth-based classification method on spike train data.  Classification on spike train data is a common task and various methods have been proposed for this purpose such as likelihood method \citep{hatsopoulos2001representations}, metric-based nearest neighbor \citep{victor1997metric, van2001novel}, minimum-distance-to-mean \citep{wu2013estimating}, and maximum-depth classifier \citep{liu1990notion}.  When a median spike train is known, one can simply adopt a minimum-distance-to-median for the classification.  In this paper, we propose a more powerful method which is based on the well-known depth-depth (DD) classifier.  The DD classifier was proposed in  \citet{li2012dd} to improve the maximum-depth classifier by finding an optimal boundary function in a depth-depth plot \citep{liu1999multivariate}.  For a 45 degree boundary line, the method is the same as the maximum-depth classifier.  We point out the DD classifier does not have an increasing boundary line, which lacks reasonable interpretation in practical use.  In this paper, we develop a modified version of the DD classifier which guarantees the monotonicity of the boundary function.  We will systematically compare the proposed classification with state-of-the-art mentioned above using simulations and real experimental data.

The rest of this paper is organized as follows. In Section \ref{sec:method}, we will first review the notion of depth for spike train and give the definition of depth-based median spike train. Simulations will be conducted to show the superiority of median spike train over mean spike train. Then, we will develop a depth-based method to detect outliers in spike train sample as well as a new supervised classification approach.  In Section \ref{sec:results}, we will use two simulations and two real experimental datasets to illustrate the effectiveness and superiority of our framework. The summary and future work will be given in Section \ref{sec:summary}.  All mathematical proofs will be shown in Appendix. 

\section{Methods} \label{sec:method}
In this section, we will provide all mathematical details in the new framework. We will at first review the notion of depth in point process and apply it to spike train data. 

\subsection{Spike train depth}
In this paper, we will adopt the framework in \citet{zhou2023statistical} to define the spike train depth. Let $\mathbb{S}$ be the set of all spike trains in a fixed time domain $[T_1,T_2]$.  Then $\mathbb{S} = \cup_{k=0}^\infty \mathbb{S}_k$, where $\mathbb{S}_k$ is the set of all spike trains with cardinality $k$ in $[T_1,T_2]$. 
The depth on spike trains is a function that maps from $\mathbb{S}$ to $\mathbb{R}^+$. By convention, depth value will be normalized as a real number in $[0,1]$. The depth value for any spike train $\bm{s}\in\mathbb{S}$ is the product of two terms \citep{zhou2023statistical}: One is a normalized depth of the cardinality of $\bm{s}$, and the other is a conditional depth for $\bm{s}$ given its cardinality. The formal definition of depth on any spike train is given as follows: 
\begin{definition} \label{def:wholedef}
{\bf (spike train depth)} 
Let $S\in\mathbb{S}$ be a random realization of a spike train on the time domain $[T_1,T_2]$ with conditional intensity function $\lambda(t\arrowvert H_t)$, where $H_t$ denotes the spike history at time $t$. Denote $P_{\arrowvert S\arrowvert}$ as a probability measure on the cardinality $\arrowvert S\arrowvert$. Then, for any spike train $\bm{s}\in \mathbb{S}$, its depth $D(\bm{s};\lambda)$ is defined as: 
\begin{eqnarray}
D(\bm{s};\lambda)=w(\arrowvert\bm{s}\arrowvert;P_{\arrowvert S\arrowvert})^rD_\lambda(\bm{s} \mid \arrowvert\bm{s}\arrowvert),
\end{eqnarray}
where $w(\arrowvert\bm{s}\arrowvert;P_{\arrowvert S\arrowvert})=\frac{D_1(\arrowvert\bm{s}\arrowvert;P_{\arrowvert S\arrowvert})}{\max_kD_1(\arrowvert\bm{s}\arrowvert=k;P_{\arrowvert S\arrowvert})}$ is the normalized depth on the cardinality $\arrowvert\bm{s}\arrowvert$ with $D_1(\arrowvert\bm{s}\arrowvert;P_{\arrowvert S\arrowvert})=\min\{P_{\arrowvert S\arrowvert}(\arrowvert S\arrowvert\leq \arrowvert\bm{s}\arrowvert),P_{\arrowvert S\arrowvert}(\arrowvert S\arrowvert\geq \arrowvert\bm{s}\arrowvert)\}$, $r>0$ is a hyper-parameter, and $D_\lambda(\bm{s} \mid \arrowvert\bm{s}\arrowvert)$ is the conditional depth of $\bm{s}$ conditioned on $\arrowvert\bm{s}\arrowvert$. 
\end{definition}

\begin{remark}
There exist multiple methods to estimate the one dimensional depth $D_1(\arrowvert\bm{s}\arrowvert;P_{\arrowvert S\arrowvert})$ in addition to the one given in Definition \ref{def:wholedef}. 
In practice, the empirical estimates of $D_1(\arrowvert\bm{s}\arrowvert;P_{\arrowvert S\arrowvert})$ and $w(\arrowvert\bm{s}\arrowvert;P_{\arrowvert S\arrowvert})$ can be applied to substitute the population result when the population distribution is unknown. 
\end{remark}

The conditional depth $D_\lambda(\bm{s} \mid \arrowvert\bm{s}\arrowvert)$ depends on the temporal distribution of the spike train times.  These times are in increasing order in a finite range and traditional multivariate methods cannot be directly used to provide a proper rank.
In this paper, we will adopt the newly developed ILR conditional depth \citep{zhou2023statistical}, and the detail will be given in next subsection. 

\subsection{The ILR depth and a simplified version}
The ILR depth is a density-based depth via the well known Isometric Log-Ratio (ILR) transformation. For any spike train $\bm{s}=(s_1,s_2,\dots,s_k)\in\mathbb{S}_k$ with cardinality $k \in \mathbb R^+$, denote $s_0=T_1$, $s_{k+1}=T_2$. Then the vector of inter-spike intervals (ISI) $\bm{u}=(s_1-s_0,s_2-s_1,\dots,s_{k+1}-s_k)$ belongs to the simplex $\mathcal{S}^{k+1} = \{(x_1, \cdots, x_{k+1}) \in \mathbb R^{k+1} \mid x_i \ge 0, i = 1, \cdots, k+1, \sum_{i=1}^{k+1} x_i = T_2-T_1\}$. Then, the ILR transformation of $\bm{u}\in\mathcal{S}^{k+1}$ is given as the following formula: $\bm{u}^*=ilr(\bm{u})  =   \Psi\cdot\Big[\log\frac{u_1}{g(\bm{u})},\log\frac{u_2}{g(\bm{u})},\dots,\log\frac{u_{k+1}}{g(\bm{u})}\Big]^T$, where $g(\bm{u})$ is the geometric mean of $\bm{u}$.  $\Psi$ is a matrix in $\mathbb R^{k\times (k+1)}$ which satisfies $\Psi\Psi^T=I_k$ and $\Psi^T\Psi=I_{k+1}-\frac{1}{k+1}\bm{1}_{k+1}\bm{1}_{k+1}^T$, where $I_k$ is the identity matrix in $\mathbb R^{k \times k}$, $I_{k+1}$ is the identity matrix in $\mathbb R^{(k+1) \times (k+1)}$, and $\bm{1}_{k+1}$ is a column vector of ones in $\mathbb R^{k+1}$. Thus, $\bm{u}^*$ is the ILR transformation result in $\mathbb R^{k}$. 

If the spike train $\bm{s}$ is a realization from a homogeneous Poisson process, then its ISI vector $\bm{u}$ is uniformly distributed in $\mathcal{S}^{k+1}$ and the transformed vector follows a simplicial distribution in $\mathbb R^{k}$ \citep{zhou2023statistical}.  This distribution is adopted to define the ILR depth.  In general, a point process can be transformed to a homogeneous Poisson process by the well-known time rescaling method \citep{brown2002time} and then the ILR depth is defined on the transformed data. Formally, the definition of the ILR depth for any spike train conditioned on cardinality $k$ is given as follows: 

\begin{definition} \label{def:ilr}
{\bf (ILR depth)} 
For point process with conditional intensity function $\lambda(t\arrowvert H_t)$  on the time domain $[T_1,T_2]$, denote $\Lambda(t)=\int_{T_1}^t\lambda(u\arrowvert H_u)du$, $s_0=T_1$ and $s_{k+1}=T_2$.  For any spike train $\bm{s}=(s_1,s_2,\dots,s_k)\in\mathbb{S}_k$, its ILR depth with respect to $\lambda(\cdot)$ conditioned on $\arrowvert\bm{s}\arrowvert=k$ is defined as: 
\begin{equation}
\begin{aligned}
    D_{\lambda-ILR}(\bm{s} \mid \arrowvert\bm{s}\arrowvert=k)
    =\frac{1}{1-\log\Big(\frac{(k+1)^{k+1}}{\Lambda(T_2)^{k+1}}\prod_{i=1}^{k+1}(\Lambda(s_i)-\Lambda(s_{i-1}))\Big)}.
\end{aligned}
    \label{eq:cdepth}
\end{equation}
\end{definition}

\begin{remark} If the spike train is a realization of a homogeneous Poisson process with constant intensity rate $\lambda$ on $[T_1,T_2]$, then its ILR depth can be simplified as: 
\begin{equation*}
\begin{aligned}
D_{\lambda-ILR}(\bm{s} \mid \arrowvert\bm{s}\arrowvert=k)
=\frac{1}{1-\log\big(\frac{(k+1)^{k+1}}{(T_2-T_1)^{k+1}}\prod_{i=1}^{k+1}(s_i-s_{i-1})\big)}.
\end{aligned}
\end{equation*}
\end{remark}

In practice, we can only have spike train observations and the conditional intensity function $\lambda(t\arrowvert H_t)$ is unknown.  
Due to the high complexity on the history condition, simplifications are often needed to estimate $\lambda(t\arrowvert H_t)$.  Poisson process is most commonly used, where the spike each time is independent of the spike history and the intensity can be easily estimated via a kernel smoothing method \citep{silverman1986density}.  For non-Poisson process,  an inhomogeneous Markov interval (IMI) method has been introduced to model firing pattern \citep{kass2001spike}.  The IMI method assumes that $\lambda(t\arrowvert H_t)$ only depends on the current time index $t$ and the last time event preceding to time $t$. This simplification is reasonable for many spike trains and can greatly reduce the estimation complexity. In the remaining part of this paper, we will adopt the IMI method to estimate $\lambda(t\arrowvert H_t)$ for non-Poisson processes. Once the estimated intensity $\hat{\lambda}(t\arrowvert H_t)$ is obtained, it can be substituted in Eqn. \eqref{eq:cdepth} to compute the conditional depth value. 

The depth value of any spike train in a given sample can be computed via Definitions \ref{def:wholedef} and \ref{def:ilr}. However, the contours of the ILR depth value is a simplex and therefore may not be easy to use in practical data.  Based on the Laplacian approximation, a simplified version of the ILR depth was introduced in \citet{zhou2023statistical}, where the contours are the commonly desired hyper-spheres. The definition of this simplified method is given below: 

\begin{definition} \label{def:opt}  
{\bf (simplified ILR depth)} 
For point process with conditional intensity function $\lambda(t\arrowvert H_t)$ on $[T_1, T_2]$, denote $s_0=T_1$, $s_{k+1}=T_2$, $\Lambda(t)=\int_{T_1}^t\lambda(u\arrowvert H_u)du$, and $g_s=\Big(\prod_{i=1}^{k+1}\big(\Lambda(s_i)-\Lambda(s_{i-1})\big)\Big)^{\frac{1}{k+1}}$. For any spike train $\bm{s}=(s_1,s_2,\dots,s_k)\in\mathbb{S}_k$, its simplified version of the ILR depth with respect to $\lambda(\cdot)$ conditioned on $\arrowvert\bm{s}\arrowvert=k$ is defined as: 
\begin{equation*}
\begin{aligned}
    D_{\lambda-SIM}(\bm{s} \mid \arrowvert\bm{s}\arrowvert=k)
    =\frac{1}{1+\frac{1}{2}\sum_{i=1}^{k+1}\Big(\log\frac{\Lambda(s_i)-\Lambda(s_{i-1})}{g_s}\Big)^2}.
\end{aligned}
\end{equation*}
\end{definition}

In this paper, we will adopt both Definitions \ref{def:ilr} and \ref{def:opt} to rank spike train data and illustrate their use in finding templates and outliers. 

\subsection{Median spike train}

Based on the ILR depth, we will introduce the notion of median spike train in this subsection.  To the best of our knowledge, this notion has not been well investigated in the literature on spike train methods.  We will study the robustness of this median and compare it with the recently introduced mean spike train \citep{wu2013estimating}.  

\subsubsection{Definition and comparison with mean spike train}
Median is a fundamental robust summary statistic to measure the centrality of a data sample. For univariate data, median is the data point located at the center, or 50\% quantile by ordering all observations. For multivariate or point process data, by adopting the notion of depth, median is the data point with the largest depth value \citep{tukey1975mathematics}.  Based on this framework, the formal definition of median for spike train is given as below: 
\begin{definition} \label{def:median}
{\bf (median spike train)} 
Suppose $\bm{S}$ is a sample of spike trains with intensity function $\lambda(\cdot)$. Given a depth function $D(\cdot;\lambda)$ that define depth value for any spike train $\bm{s}\in\mathbb{S}$ with respect to $\bm{S}$, the median of spike train is $$Med(\bm{S})=\arg\max_{\bm{s}\in\mathbb{S}}D(\bm{s};\lambda).$$ 
\end{definition}

\begin{remark}
Based on Definition \ref{def:median}, the median spike train $\bm{s}$ does not necessarily belong to the spike train sample $\bm{S}$. In other words, the depth-based median is searched in the entire space $\mathbb{S}$. 
For a sample of HPP, the median will be close to the uniformly located spike train with maximal normalized depth $w(\cdot)$ in Definition \ref{def:wholedef}. For a sample of inhomogeneous Poisson process (IPP for short) or non-Poisson process, the median needs to be determined via the inverse of function $\Lambda(\cdot)$ in Definitions \ref{def:ilr} and \ref{def:opt}. 
\end{remark}

We will now compare the median spike train with the mean spike train \citep{wu2013estimating} in terms of robustness and efficiency.  Similar with the mean computation of Euclidean points, the mean of spike train sample $\bm{S}=\{\bm{s}_1,\bm{s}_2,\dots,\bm{s}_n\}$ is defined as 
\begin{equation}
Mean(\bm{S})=\arg\min_{\bm{s}\in\mathbb{S}}\sum_{i=1}^n d_\mu(\bm{s},\bm{s}_i)^2.
\label{eq:mean}
\end{equation}
Here $d_\mu(\cdot,\cdot)$ is a penalized metric distance with coefficient $\mu (\ge 0)$ between two spike trains which owns Euclidean properties (similar to the conventional $\mathbb L^2$ distance).  Specifically, for two spike trains $f = (t_1, \cdots, t_M)$ and $g = (s_1, \cdots, s_N)$,
\begin{equation*}
\begin{aligned}
d_\mu(f,g) = \min_\gamma \Bigg(M+N-2\sum_{i=1}^M\sum_{j=1}^N 1_{\{t_i=\gamma(s_j)\}} 
 + \mu \int_{T_1}^{T_2} (1-\sqrt{\dot \gamma})^2 dt \Bigg)^{1/2},
\end{aligned}
\end{equation*}
where $1_{\{\cdot\}}$ is an indicator function and $\gamma$ is any time warping on $[T_1, T_2]$ which satisfies $\gamma(T_1) = T_1, \gamma(T_2) = T_2,$ and the derivative $\dot \gamma > 0$.  
 
Since the mean in Eqn. \eqref{eq:mean} is determined by the sum of squares with the metric, an outlier may cause a dramatic increase to the sum and make the mean estimation unstable. In contrast, the median spike train is the deepest one in Definition \ref{def:median}.  Outliers can only slightly affect the estimation of conditional intensity function and the deepest point is relatively robust.  That is, from the definitions, median spike train is more robust to outliers compared with mean spike train. 


In terms of efficiency, the estimation of mean spike train relies on a recursive procedure to minimize the sum of squares in Eqn. \eqref{eq:mean} \citep{wu2013estimating}.  In each step, a dynamic programming procedure is needed to estimate between each sample train and the current mean.  Therefore, the estimation may be computationally expensive.  In contrast, the estimation of median spike train mainly depends on the estimation of conditional intensity function $\lambda(\cdot)$.  Once it is known, an efficient Newton-Raphson procedure will produce the median spike train.  For Poisson processes, $\lambda(\cdot)$ is deterministic and can be efficiently estimated via conventional kernel methods.  In the general point process case, $\lambda(\cdot)$ depends on history and simplification is often necessary, such as using an IMI model, to make the estimation process tractable. 

In the next subsection, we will conduct simulations to illustrate the superiority of the median over the mean in robustness and efficiency for spike trains following Poisson process model.   Comparison with more complicated non-Poisson point process will be given in Section \ref{sec:results}. 

\subsubsection{Illustrations} \label{sec:com}
In this subsection, we will simulate two spike train samples.  One is from HPP, and the other is from IPP. 

\begin{figure*} [h!]
	\centering	
	\subfigure[median, ILR depth]{
		\begin{minipage}[b]{0.3\textwidth}
			\centering
			\includegraphics[scale=0.25]{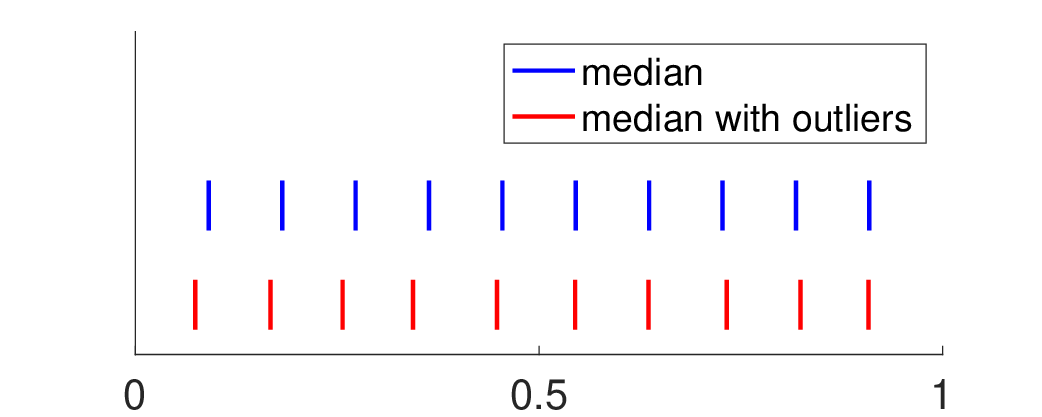}
		\end{minipage}
	}
	\subfigure[median, simplified depth]{
		\begin{minipage}[b]{0.3\textwidth}
			\centering
			\includegraphics[scale=0.25]{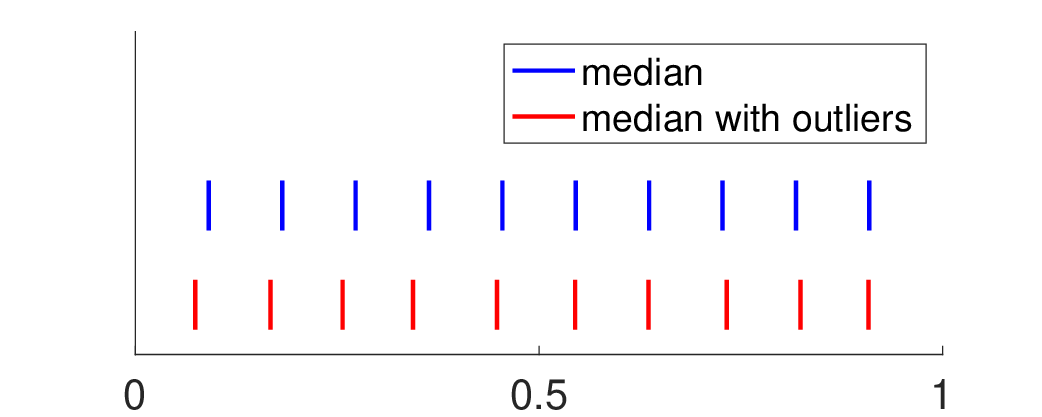}
		\end{minipage}
	}
	\subfigure[mean]{
    		\begin{minipage}[b]{0.3\textwidth}
			\centering
   		 	\includegraphics[scale=0.25]{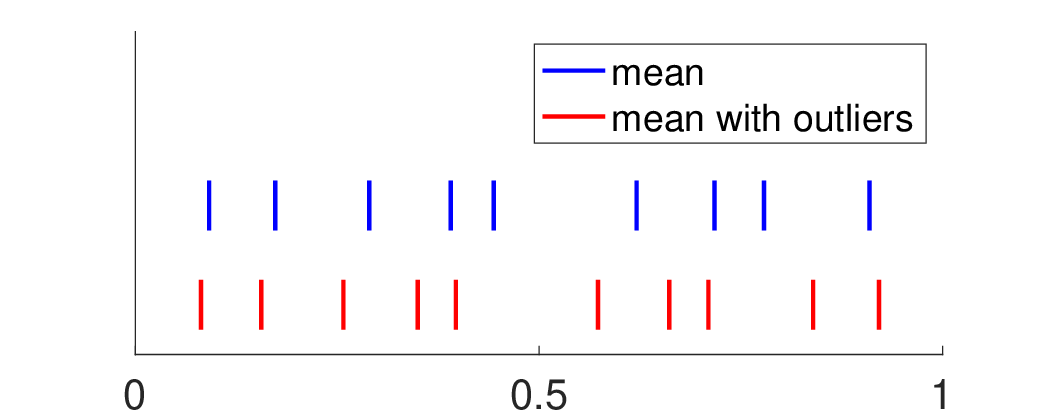}
    		\end{minipage}
    	}
	\caption{Estimation of mean and median in the HPP simulation. (a) The estimated median in the original data by using the ILR depth is shown with blue vertical lines, which indicate spike event times.  The estimated median in the outlier-included data by using the ILR depth is shown with red vertical lines. (b) Same as (a) except for using the simplified ILR depth.  (c) The estimated mean in the original data is shown with blue vertical lines.  The estimated mean in the outlier-included data is shown with red vertical lines.
}
	\label{fig:com_hpp_out}
\end{figure*}

{\bf Simulation 1:} Consider a collection of independent realizations from an HPP with sample size $n=500$. Let $T_1=0$, $T_2=1$, and the constant intensity $\lambda=10$.   The hyper-parameter $r$ in Definition \ref{def:wholedef} is taken as $1$, and we adopt both ILR depth and its simplified version to estimate the median.  The two estimated median spike trains are shown as blue vertical lines in Fig. \ref{fig:com_hpp_out}(a) and (b), respectively.  We can see that in either case, there is 10 spikes in the median and they are evenly distributed, which properly characterize the typical pattern in the homogeneous Poisson process and match the expected number of spikes (i.e., 10) in the time domain. 
For the mean spike train calculation, the penalty coefficient $\mu$ in Eqn. \eqref{eq:mean} is taken as 20.  The estimated mean by the MCP algorithm \citep{wu2013estimating} is also shown as blue vertical lines in Fig. \ref{fig:com_hpp_out}(c).  We can see that the spikes in the mean also evenly distributed in the time domain, whereas the number of spikes is 9 in this estimation, different from the expected value of 10.


To evaluate the robustness of the median and mean spike trains, 10 independent outliers are generated and added to the original simulation with new sample size $510$.  In particular, the outliers are realizations of an HPP on $[0,0.05]$ with constant intensity rate $\lambda'=200$. That is, the expected number of events in outlier process is still 10. The estimation result is also shown in Fig. \ref{fig:com_hpp_out} in red color. From Panels (a) and (b), we can see the median with either depth function changes slightly when outliers are included in the sample.  The number of spikes is still 10 and the spike times are very close to those in the original data.
In contrast, according to Fig. \ref{fig:com_hpp_out}(c), there are noteworthy distinctions between the means with or without outliers. When outliers are included, many spike times in the estimated mean are shifted toward left side from those estimated in the original data.  This coincides with the outlier type since the time events in outlier spike trains concentrate in the left side sub-interval $[0, 0.05]$.  In addition, the cardinality of mean with outliers becomes 10 rather than 9. This indicates that the cardinality of mean may be sensitive to outliers.

\begin{figure*} [h!]
	\centering
	\subfigure[median, ILR depth]{
		\begin{minipage}[b]{0.3\textwidth}
			\centering
			\includegraphics[scale=0.25]{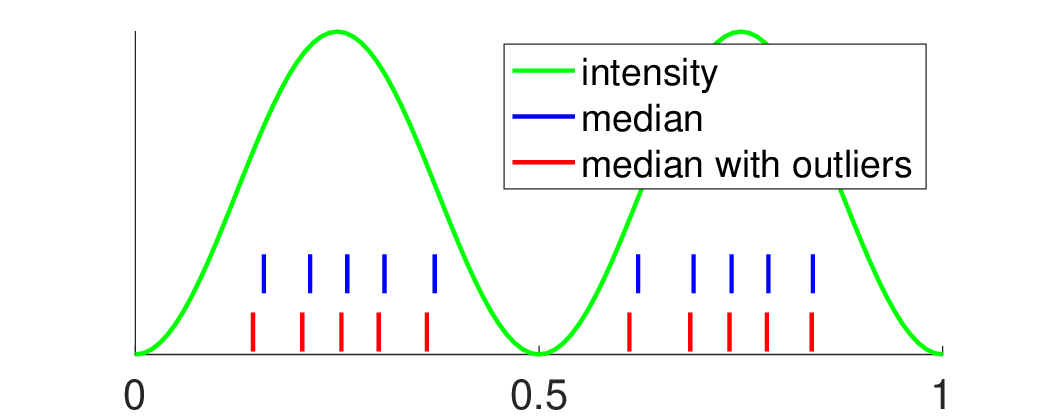}
		\end{minipage}
	}
	\subfigure[median, simplified depth]{
    		\begin{minipage}[b]{0.3\textwidth}
			\centering
   		 	\includegraphics[scale=0.25]{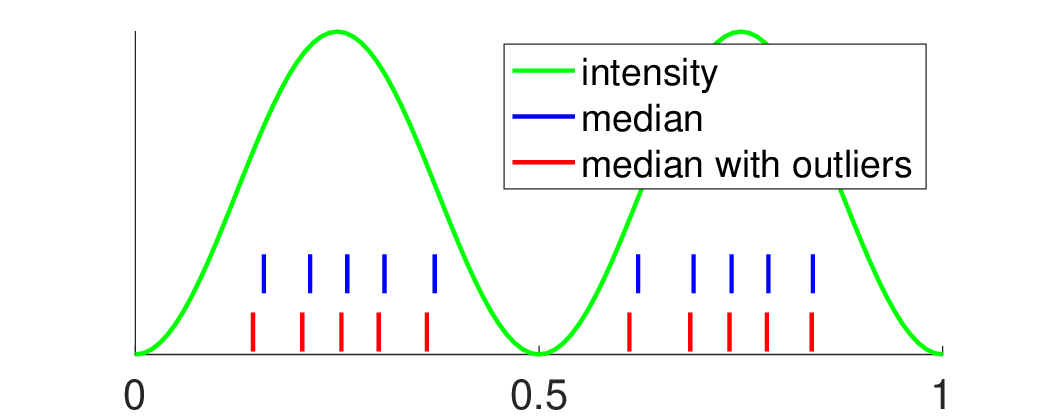}
    		\end{minipage}
    	}
	\subfigure[mean]{
		\begin{minipage}[b]{0.3\textwidth}
			\centering
			\includegraphics[scale=0.25]{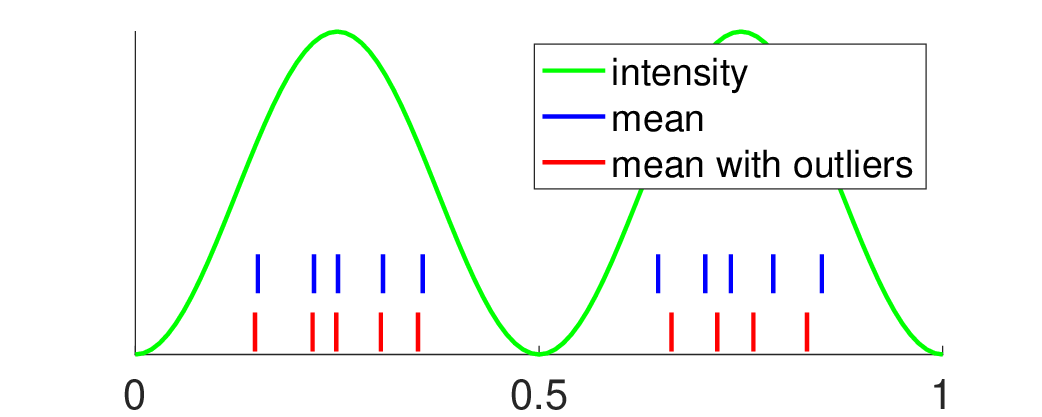}
		\end{minipage}
	}
	\caption{Same as in Fig. \ref{fig:com_hpp_out} except for the IPP simulation, where the intensity function is shown as green curve in each plot.}
	\label{fig:com_ipp_out}
\end{figure*}

{\bf Simulation 2:} Consider a sample of independent IPP with sample size $n=500$. Let $T_1=0$, $T_2=1$, the intensity function $\lambda(t)=10\sin\big(4\pi(t-\frac{1}{8})\big)+10$, and the hyper-parameter $r$ in Definition \ref{def:wholedef} be $1$. The median estimation result with the ILR depth and its simplified version are shown in Fig. \ref{fig:com_ipp_out}(a) and (b), respectively, with blue color. One can find that in either case the events are properly located around the two peaks of the intensity function. For the mean computation, the penalty term $\mu$ is set as $20$, and the result is shown in Fig. \ref{fig:com_ipp_out}(c) with blue color. Similar with the two medians, the events in the mean are distributed by following the intensity function. 

Same as in Simulation 1, 10 HPP in $[0,0.05]$ with intensity function $\lambda'=200$ are generated as outliers to the original data. The estimation results of mean and median with outliers are shown in Fig. \ref{fig:com_ipp_out}.  One can find that the median estimates are robust with respect to the outliers.  The number of spikes remain the same, and their locations only slightly change.  In contrast, the number of spikes in the estimated mean varies from 10 to 9 when the outliers are added.  This again indicates that the newly defined median spike train exhibits superior robustness to the mean spike train. 

We have compared the robustness of the mean and median using two simulations.  Now we compare the computational efficiency for their estimations. We repeat both simulations 10 times for comparison, where the device is MacBook Pro M1 2020. The average computational time for the mean is $108.88$ and $71.18$ seconds in Simulations 1 and 2, respectively. In contrast, the average computational time for the median is only $0.04$ and $0.03$ seconds, respectively. This result clearly shows the superior efficiency of the median spike train.  

In the following part of this section, we will examine two applications, outlier detection and classification, to spike train sample via the proposed spike train depth. Both studies explore novel analysis methods on spike train data.

\subsection{Outlier detection} \label{sec:out_det}
In this subsection, we will utilize the depth on spike trains to conduct outlier detection. 
To the best of our knowledge, our study in this paper is the first depth-based method on outlier detection for point process spike trains.  

\subsubsection{Definition}
We here propose to present a novel outlier detection for spike train data. 
Our framework aims to find a threshold on the depth value for each cardinality.  That is, for any observed spike train with $k$ spikes, if its depth value is less than a value $t_k$, it will be considered as a potential outlier. This idea is formally given in the following definition. 
\begin{definition} \label{def:outlier}
Given a sample of spike trains $\bm{S}$ with intensity function $\lambda(\cdot)$, for any pre-specified fixed number $\delta\in(0,1)$, a spike train $\bm{s}_0$ with cardinality $k$ is a potential outlier of $\bm{S}$ if $D(\bm{s}_0;\lambda)<t_k$, where $t_k$ satisfies
\begin{eqnarray} \label{eqn:out_prob}
P\big(D(\bm{s}_0;\lambda)<t_k \big)=\delta.
\end{eqnarray}
\end{definition}

\begin{remark}
The parameter $\delta$ in Definition \ref{def:outlier} is a hyper-parameter, its proper value may vary with respect to different samples. When the intensity $\lambda(\cdot)$ is unknown, it can be substituted by the estimated intensity $\hat{\lambda}(\cdot)$ from the given sample $\bm{S}$. 
\end{remark}

In this paper, we focus on finding a proper value of $t_k$ for each cardinality $k$ in a given spike train sample. We will at first study the threshold value for Poisson process, and then generalize the method to general point process.

\subsubsection{Threshold in Poisson process}
\label{sec:th_poission}
For Poisson process, the intensity function is deterministic and history independent. Thus, the threshold value $t_k$ in Eqn. \eqref{eqn:out_prob} can be easily derived by using the ILR depth. 
We state the main result as follows, where the proof is given in Appendix \ref{app:outlier}. 
\begin{proposition} \label{prop:tk}
Given a positive number $\delta$ in Definition \ref{def:outlier} and a sample of Poisson process in $[T_1,T_2]$ with intensity function $\lambda(t)$, denote $\Lambda(t)=\int_{T_1}^{t}\lambda(s)ds$. Suppose $\bm{U}=(U_1,U_2,\dots,U_k)$ is a vector of $k$ ordered uniform random variable in $[0,\Lambda(T_2)]$, and denote $U_0=0$, $U_{k+1}=\Lambda(T_2)$. Let $G=\prod_{i=1}^{k+1}(U_i-U_{i-1})$ and $C_k=F_G^{-1}(\delta)$, where $F_G(\cdot)$ is the cumulative density function of $G$. By using the ILR depth, the threshold $t_k$ for any cardinality $k (\ge 0)$ is given as
\begin{eqnarray}
t_k=\frac{w(k;P_{\arrowvert S\arrowvert})^r}{1-\log\bigg(C_k\cdot \Big(\frac{k+1}{\Lambda(T_2)}\Big)^{k+1}\bigg)} \label{eq:tk}.
\end{eqnarray}
\end{proposition}

We will now use two simulations to evaluate the performance of this outlier detection framework.  The simulation procedures are very similar to those in Section \ref{sec:com}. 

{\bf Simulation 3:} We generate 1000 independent spike trains from an HPP on $[0, 1]$ with constant intensity rate $\lambda=10$ as original data. We then generate 10 HPP realizations as outliers with constant intensity rate $\lambda'=100$ on time domains $[0,0.1],[0.1,0.2],\dots,$ and $[0.9,1]$, respectively.  
 
{\bf Simulation 4:} We generate 1000 independent spike trains from an IPP on $[0,1]$ with intensity function $\lambda(t)=10\sin\big(4\pi(t-\frac{1}{8})\big)+10$ as original data. Then, 10 outliers are generated in the same way as in Simulation 3. 

\begin{figure*} [h!]
	\centering
	\subfigure[10 HPP realizations]{
		\begin{minipage}[b]{0.3\textwidth}
			\centering
			\includegraphics[scale=0.25]{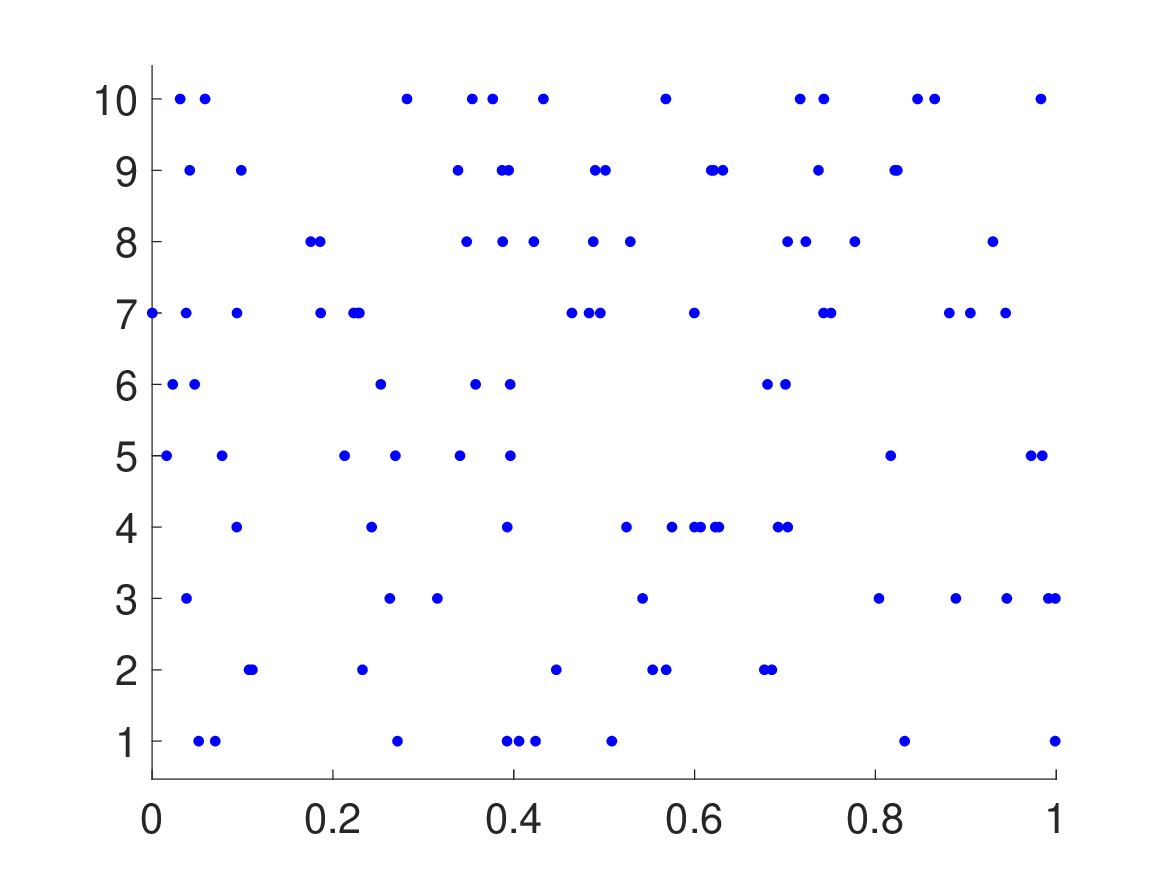}
		\end{minipage}
	}
	\subfigure[10 IPP realizations]{
		\begin{minipage}[b]{0.3\textwidth}
			\centering
			\includegraphics[scale=0.25]{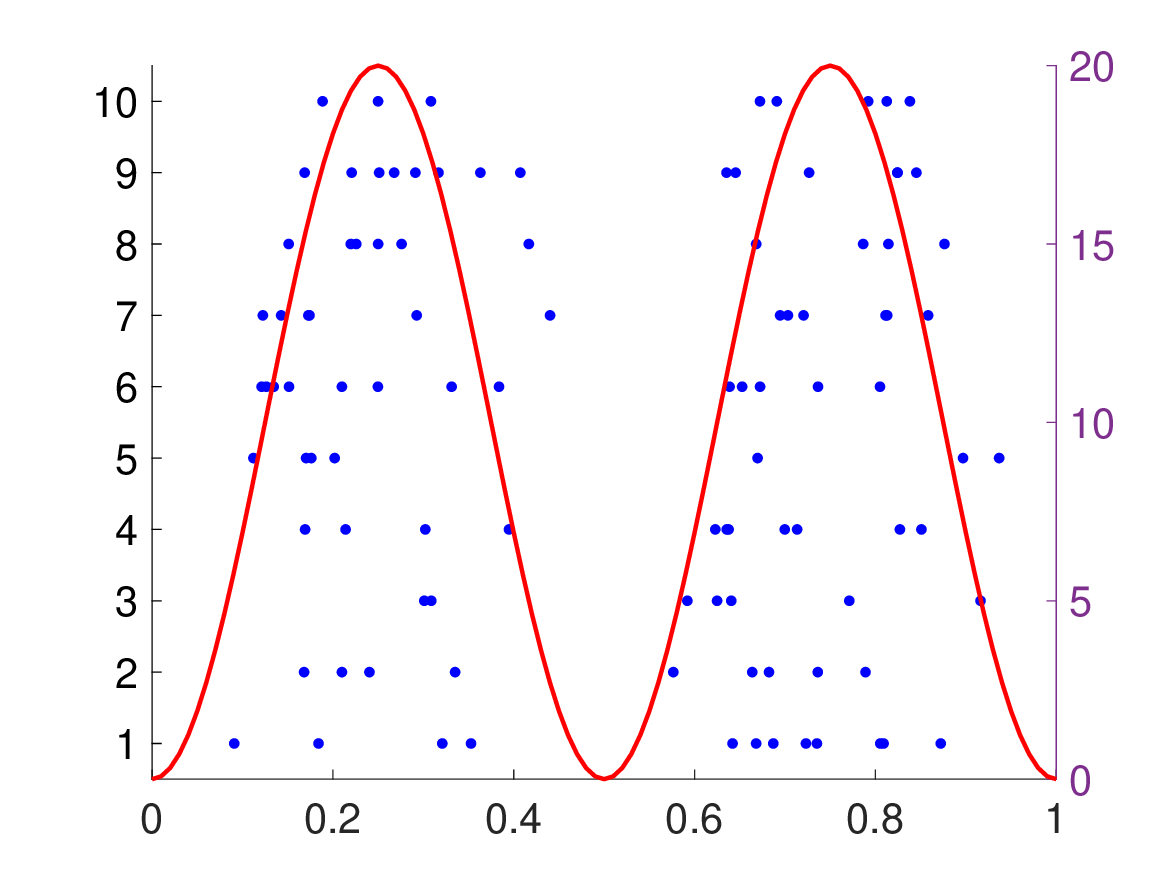}
		\end{minipage}
	}
	\subfigure[outliers]{
    		\begin{minipage}[b]{0.3\textwidth}
			\centering
   		 	\includegraphics[scale=0.25]{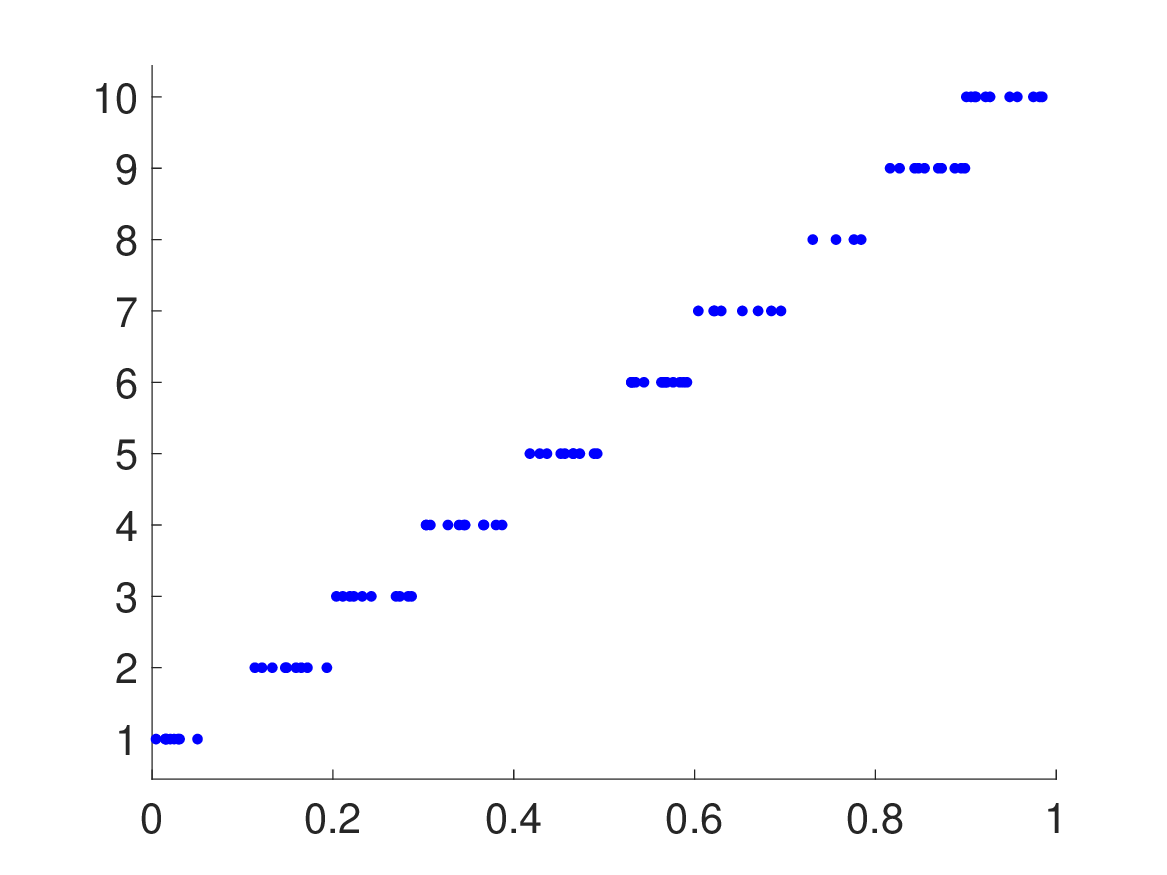}
    		\end{minipage}
    	}
	\caption{Original spike train examples and outliers. (a) 10 example original spike trains in Simulation 3, where each row represents one realization, and the spike events are marked as blue dots, the y-axis shows the indices of spike train. (b) Same as (a) except for 10 example spike trains in Simulation 4, where the red curve represents the intensity function $\lambda(t)$ and its value is shown along the right y-axis. (c) Same as (a) except for the 10 outliers in both Simulations 3 and 4. }
	\label{fig:outlier_show}
\end{figure*}

Some original realizations and outlier spike trains are shown in Fig. \ref{fig:outlier_show}. For each simulation setting, we will repeat the experiments for 100 times and summarize three quantities to evaluate the outlier detection performance. 
\begin{itemize}
\item Precision: the ratio (in percentage) between correctly detected outliers and total outliers detected. 
\item Recall: the ratio (in percentage) between correctly detected outliers and total outlier number. 
\item $F_1$ score: $\frac{2}{\text{recall}^{-1}+\text{precision}^{-1}}$.
\end{itemize}
Several values of hyper-parameter $\delta$ in Definition \ref{def:outlier} are used and the averaged result (by using mean and standard deviation) over 100 repetitions is shown in Table \ref{tab:out_com}. As comparison, the detection results of the 3S method introduced in \citet{shchur2021detecting} are shown in Table. \ref{tab:out_com}.  
\begin{table*}[h!]
\centering
\begin{tabular}{|c|c|c|c|c|}
\hline
\textbf{Data} & \textbf{method} & \textbf{precision} & \textbf{recall} & \textbf{$F_1$ score} \\ \hline
\multirow{6}{*}{Simulation 3} &  Depth: $\delta=$ 0.001  & 87.9 (0.10) & 86.1 (0.13) & \textbf{86.3 (0.09)} \\ 
                                        &    Depth: $\delta=$ 0.005 & 65.9 (0.09) & 94.1 (0.07) & \textbf{77.0 (0.07)} \\ 
                                        &     Depth: $\delta=$ 0.01 & 50.7 (0.09) & 95.8 (0.06) & \textbf{65.9 (0.08)} \\  
                                         & $3S$: threshold = 0.01 &  40.9 (0.10) & 60.9 (0.11) & \textbf{47.9 (0.08)} \\ 
                                         & $3S$: threshold = 0.03 &  25.2 (0.04) & 95.0 (0.08) & \textbf{39.5 (0.05)} \\ 
                                        & $3S$: threshold = 0.05 &  16.9 (0.03) & 99.9 (0.01) & \textbf{28.8 (0.04)} \\ \hline
\multirow{6}{*}{Simulation 4} &  Depth: $\delta=$ 0.001  & 88.3 (0.10) & 81.1 (0.11)  & \textbf{84.1 (0.09)} \\ 
                                        &    Depth: $\delta=$ 0.005 & 65.5 (0.11) & 89.8 (0.08) & \textbf{75.2 (0.08)} \\ 
                                        &     Depth: $\delta=$ 0.01 & 49.3 (0.08) & 93.5 (0.07) & \textbf{64.2 (0.07)} \\ 
                                         & $3S$: threshold = 0.01 &  38.6 (0.11) & 48.6 (0.14) & \textbf{41.5 (0.08)} \\ 
                                         & $3S$: threshold = 0.03 &  24.8 (0.04) & 93.0 (0.08) & \textbf{39.0 (0.05)} \\ 
                                        & $3S$: threshold = 0.05 &  17.4 (0.04) & 99.8 (0.01) & \textbf{29.5 (0.05)} \\ \hline

\end{tabular}
\caption{Outlier detection performance with different methods in both simulations.  The mean of all 100 trials is shown in each cell, with standard deviation shown in parentheses.  }
\label{tab:out_com}
\end{table*}
From this table, one can find the $F_1$ score of our new method can achieve around $85\%$ for both simulations. Thus, this new outlier detection framework is highly effective. The overall performance in Simulation 3 is better than that in Simulation 4. As the intensity function has peaks in Simulation 4, there exists cluster phenomena among non-outlier realizations, which makes the detection method less sensitive to outliers. In addition, as $\delta$ increases, precision tends to decrease, whereas recall tends to increase.  
This is a common tradeoff between precision and recall, and adjustment of the $\delta$ value is necessary for an effective detection. Finally, comparing our framework with the  3S method, the $F_1$ score clearly demonstrates the superiority of our framework. Although the recall remains at a roughly same level, the precision in our method is much higher. 

\subsubsection{Threshold in general point process}  
In this subsection, we will extend the proposed outlier detection framework in Sec. \ref{sec:th_poission} to non-Poisson spike trains. In this case, the conditional intensity function depends on the history of time events and the $\Lambda(T_2)$ in Eqn. \eqref{eq:tk} will change for each spike train observation. Therefore, for a sample of non-Poisson spike trains, the function $\Lambda(\cdot)$ needs to be evaluated for each observation as well as the corresponding $t_k$. 
Fortunately, this procedure is still feasible, albeit with higher computational cost. We will show one simulation as follows. 

\begin{figure*} [h!]
	\centering
	\subfigure[10 original realizations of Hawkes process]{
		\begin{minipage}[b]{0.47\textwidth}
			\centering
			\includegraphics[scale=0.3]{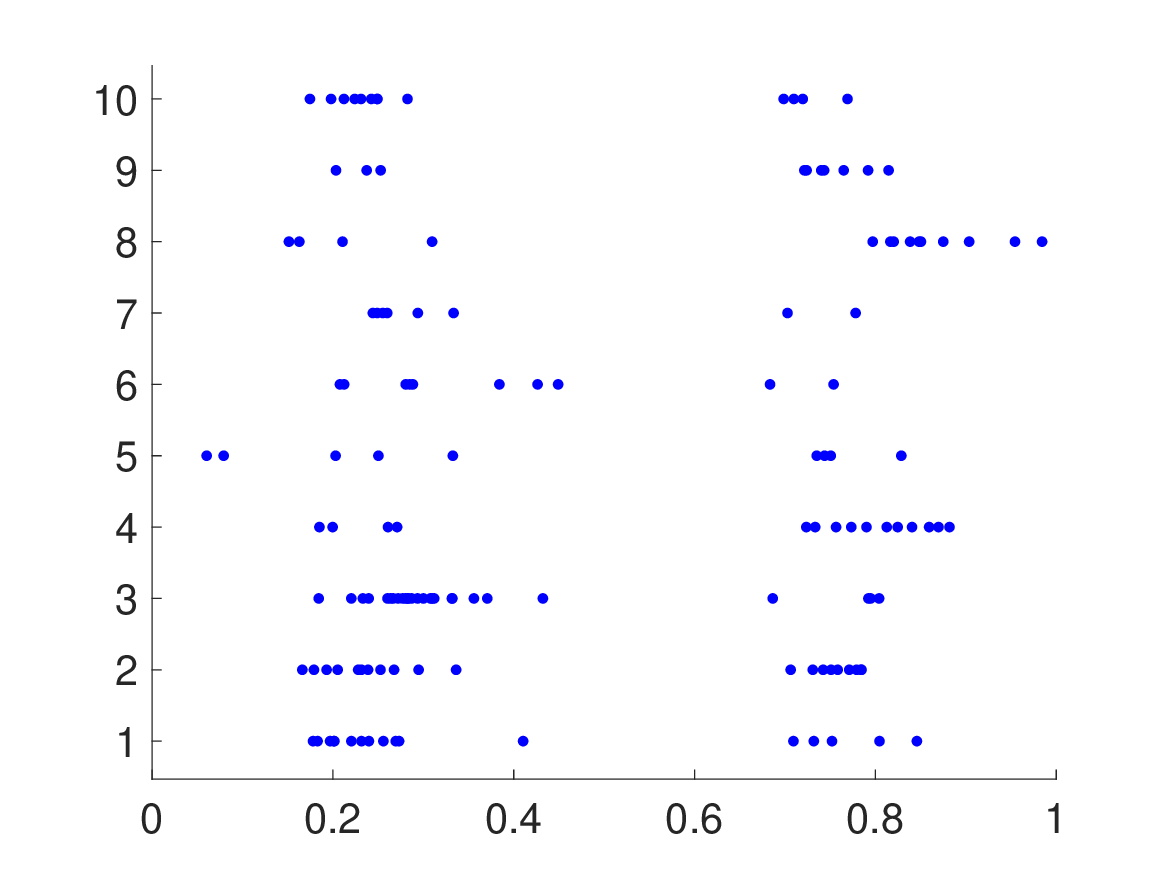}
		\end{minipage}
	}
	\subfigure[outliers]{
		\begin{minipage}[b]{0.47\textwidth}
			\centering
			\includegraphics[scale=0.3]{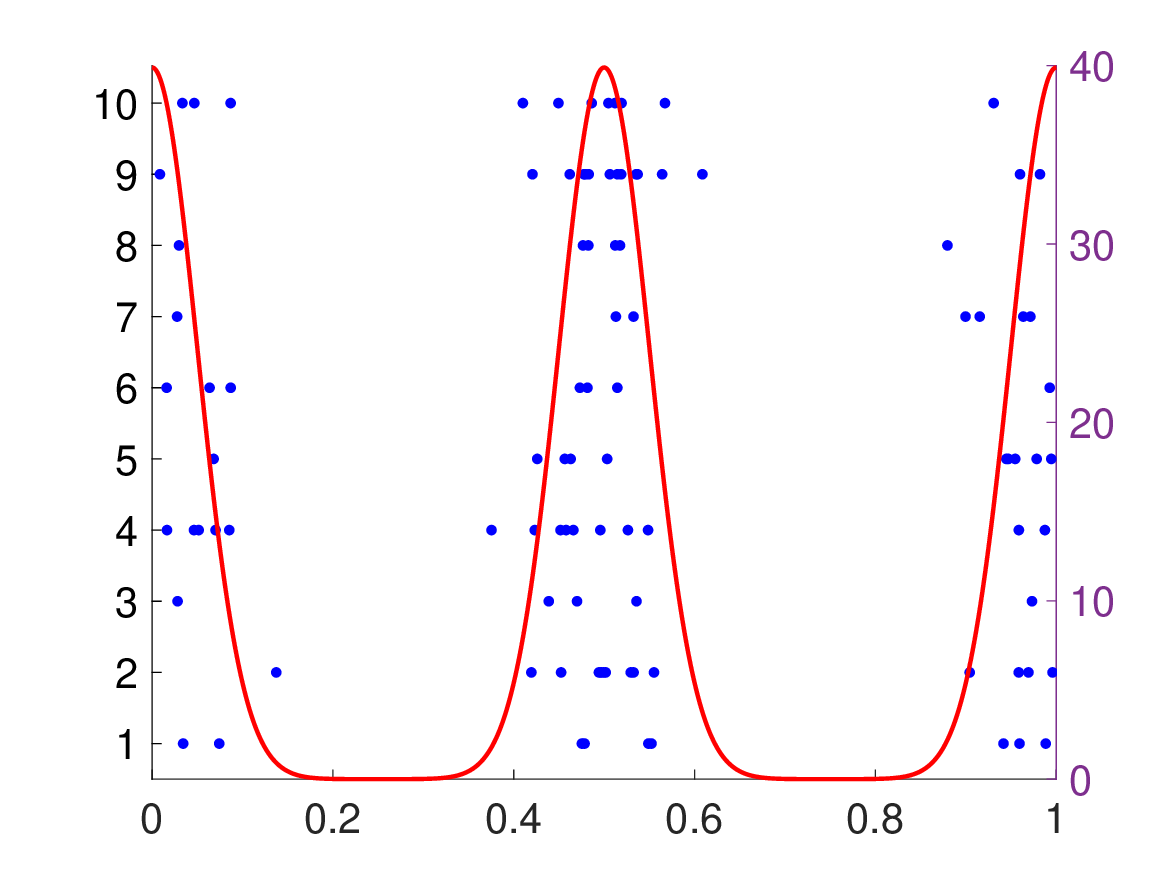}
		\end{minipage}
	}
	\caption{Original realizations and outlier spike train observations. (a) 10 example original realizations of Hawkes process in Simulation 5, where each row represents one realization and the spike events are marked as blue dots. (b) Same as (a) except for the 10 outliers in Simulation 5, where the red curve represents the intensity function $\gamma(t)$ and its value is shown along the right y-axis. }
	\label{fig:hawkes_outlier}
\end{figure*}

{\bf Simulation 5:} Generate 1000 independent spike trains as original data from a Hawkes process on $[0, 1]$ with conditional intensity function $\lambda^*(t\arrowvert H_t)=\frac{1}{2}\lambda(t)+\sum_{t_i<t}\alpha e^{-\beta(t-t_i)}$, where $\lambda(t)=\frac{100}{\sqrt{2\pi}}e^{-\frac{(t-0.25)^2}{2\times 0.05^2}}I(t\leq 0.5)+\frac{100}{\sqrt{2\pi}}e^{-\frac{(t-0.75)^2}{2\times 0.05^2}}I(t>0.5)$. 
$I(\cdot)$ is the indicator function, $\alpha=15$, $\beta=30$, $t_i$ is the time events before time $t$, and $H_t$ represents the history of time events up to time $t$. Some realizations are shown in Fig. \ref{fig:hawkes_outlier}(a). 
One can find the spike trains mainly cluster around two time regions near 0.25 and 0.75, whereas there are few spikes located at the left-most, middle or right-most region. Thus, to make the outliers distinguishable, 10 IPP are generated as outliers with intensity function $\gamma(t)=\frac{100}{\sqrt{2\pi}}e^{-\frac{t^2}{2\times 0.05^2}}I(t\leq 0.25)+\frac{100}{\sqrt{2\pi}}e^{-\frac{(t-0.5)^2}{2\times 0.05^2}}I(0.25<t\leq 0.75)+\frac{100}{\sqrt{2\pi}}e^{-\frac{(t-1)^2}{2\times 0.05^2}}I(t>0.75)$, $I(\cdot)$ is the indicator function. The outliers are shown in Fig. \ref{fig:hawkes_outlier}(b), and the events are located at the left-most, middle and right-most region.  

Same as in Simulations 3 and 4, the outlier detection procedure will be repeated 100 times and the average result is shown in Table \ref{tab:out_hawkes}. 
\begin{table*}[h!]
\centering
\begin{tabular}{|c|c|c|c|c|}
\hline
\textbf{Data} & \textbf{method} & \textbf{precision} & \textbf{recall} & \textbf{$F_1$ score} \\ \hline
\multirow{6}{*}{Simulation 5} &  Depth: $\delta=$ 0.001  & 80.8 (0.12) & 79.3 (0.13)  & \textbf{79.2 (0.10)} \\ 
                                        &  Depth: $\delta=$ 0.005 & 56.4 (0.12) & 88.9 (0.10) & \textbf{68.4 (0.10)} \\ 
                                        &  Depth: $\delta=$ 0.01 & 40.7 (0.09) & 90.6 (0.10) & \textbf{55.6 (0.08)} \\ 
                                        & $3S$: threshold = 0.05 & 8.3 (0.03) & 46.2 (0.15) & \textbf{14.0 (0.05)} \\ 
                                        & $3S$: threshold = 0.03 & 8.5 (0.04) & 29.0 (0.14) & \textbf{13.1 (0.06)} \\ 
                                        & $3S$: threshold = 0.01 & 9.0 (0.08) & 9.4 (0.09) & \textbf{8.4 (0.08)} \\ \hline
\end{tabular}
\caption{Outlier detection performance.  Same as those in Table \ref{tab:out_com} except for the Hawkes process simulation. }
\label{tab:out_hawkes}
\end{table*}
The best $F_1$ score is close to $80\%$ for this simulation. Based on the $F_1$ score, one can conclude the result again shows an effective performance of the proposed outlier detection method for non-Poisson spike trains. In comparison, the performance of the 3S method is very poor with respect to different threshold values. 

\subsection{Classification} \label{sec:classification}
In this subsection, we will adopt the main idea of depth-depth (DD) plot method \citep{li2012dd} to classify different spike train samples. Furthermore, we will conduct a new algorithm to find a strictly increasing function to define the boundary, which can make the boundary more interpretable.

\subsubsection{Framework and algorithm}
Suppose there are two groups of spike trains $ F = \{\bm{x}_i\}_{i=1}^m$ and $ G = \{\bm{y}_i\}_{i=1}^n$  with sample sizes $m$ and $n$, respectively.
 Then, for any $\bm{s}\in F\cup G$, denote $D_F(\bm{s})$ as the depth value of $\bm{s}$ with respect to group $F$, and $D_G(\bm{s})$ as the depth value of $\bm{s}$ with respect to group $G$. The DD plot \citep{liu1999multivariate} is defined as the scatter plot of $DD(F,G)=\{(D_F(\bm{s}),D_G(\bm{s})),\bm{s}\in F\cup G\}$. 

The DD classifier aims to find an optimal strictly increasing function $f(\cdot)$ satisfying $f(0) = 0$ in the DD plot such that the two groups data can be best separated. For a new spike train $\bm{s}'$, it will be assigned to group F if $f(D_F(\bm{s}'))>D_G(\bm{s}')$, or group G if $f(D_F(\bm{s}'))<D_G(\bm{s}')$. In this paper, we adopt the same idea: Assume $f(\cdot)$ is a strictly increasing function with $f(0)=0$. Given the boundary function $f(\cdot)$, according to \citet{li2012dd}, the misclassification rate $M(f)$ can be defined as: 
\begin{equation} \label{eq:mis}
\begin{aligned}
M(f)=\frac{1}{m+n}\Big(\sum_{i=1}^{m}I\big(D_G(\bm{x}_i)>f(D_F(\bm{x}_i))\big)  
 +\sum_{i=1}^{n}I\big(D_G(\bm{y}_i)<f(D_F(\bm{y}_i))\big)\Big),
\end{aligned}
\end{equation}
where $I(\cdot)$ is the indicator function and the optimal solution of $f$ is obtained if $M(f)$ in Eqn. \eqref{eq:mis} is minimized.

In this paper, we propose a new algorithm to obtain $f(\cdot)$ such that it is a strictly increasing function. By this assumption,  $h(x)=\log(\dot{f}(x))$ will be an unconstrained function in the conventional Euclidean space. It is straightforward to see $f$ can also be represented by $h$ as $f(t)=\int_{0}^{t}e^{h(x)}dx$. Thus, Eqn. \eqref{eq:mis} can be rewritten as: 
\begin{eqnarray} \label{eq:mish}
M(h)&=&  \frac{1}{m+n}\Big(\sum_{i=1}^{m}I\big(D_G(\bm{x}_i)>\int_{0}^{D_F(\bm{x}_i)}e^{h(x)}dx\big)  \nonumber \\
  & & \quad\quad\quad\quad + \sum_{i=1}^{n}I\big(D_G(\bm{y}_i)<\int_{0}^{D_F(\bm{y}_i)}e^{h(x)}dx\big)\Big).
\end{eqnarray}
By the equivalent representation between $f$ and $h$, our goal is to find an optimal $h(\cdot)$ such that $M(h)$ in Eqn. \eqref{eq:mish} is minimized. 
To make the optimization process feasible, we assume $h(\cdot)$ is a polynomial with maximal degree $k_0$, e.g. $h(x)=\sum_{i=0}^{k_0}a_i x^{i}$. In this way, the optimization becomes a parametric problem and thereby can be solved by gradient methods. Since the indicator function is not differentiable, one can use logistic function to approximate it, i.e., $I(x>0)\approx \frac{1}{1+e^{-tx}}$ for an appropriate positive value of $t$. In this paper, we will assume $t=100$, the value suggested by \citet{li2012dd}, to train the DD classifier. The optimization procedure is summarized in Algorithm \ref{alg:dd_opt} as follows. 
\begin{algorithm}[ht!]
\caption{Stochastic gradient method for the DD classifier}
\begin{algorithmic}
\State{\textbf{Input}: Four groups of depth value $D_F(\bm{x}_i)$, $D_G(\bm{x}_i)$, $D_F(\bm{y}_j)$, $D_G(\bm{y}_j)$ for $i=1,2,\dots,m$ and $j=1,2,\dots,n$; Misclassification rate function $M$; Maximal degree $k_0$; Initial coefficient guess $\bm{a}=[a_{k_0},a_{k_0-1},\dots,a_1,a_0]^T$; Learning rate $\delta$, Annealing parameter $\alpha$; stopping criterion $\epsilon$.}
\State{- Generate $Z\sim N(\bm{0},I_{k_0+1})$, $\bm{0}$ is $(k_0+1)$-dim column vector with all entries $0$, $I_{k_0+1}$ is $(k_0+1)\times(k_0+1)$ identity matrix;}
\State{- Denote $T=1$;}
\State{- Compute $\bm{a}_{new}=\bm{a}-\delta\nabla_{\bm{a}}M+\sqrt{\delta T}\cdot Z$;}
\While{$\| \bm{a}_{new}-\bm{a}\|>=\epsilon$}
	\State{- $T=\alpha T$;}
	\State{- $\bm{a}=\bm{a}_{new}$;}
	\State{- $\bm{a}_{new}=\bm{a}-\delta\nabla_{\bm{a}}M+\sqrt{\delta T}\cdot Z$;}
\EndWhile
\State{- Compute $f(t)=\int_{0}^{t}e^{[x^{k_0},x^{k_0-1},\dots,1,0]\bm{a}_{new}}dx$ for $t\in[0,1]$;}
\State{\textbf{Output}: Optimal boundary function $f(t)$.}

\end{algorithmic}
\label{alg:dd_opt}
\end{algorithm}


Algorithm \ref{alg:dd_opt} will return the boundary function of DD classifier, and the classification can be conducted based on the function output. Moreover, similar to all depth-based classification methods, Algorithm \ref{alg:dd_opt} can be applied to classify any object with well defined depth. 

\subsubsection{Illustration with multivariate data} \label{sec:cl_illus}
\begin{figure*} [h!]
	\centering
	\subfigure[DD plot with boundary functions]{
		\begin{minipage}[b]{0.47\textwidth}
			\centering
			\includegraphics[scale=0.3]{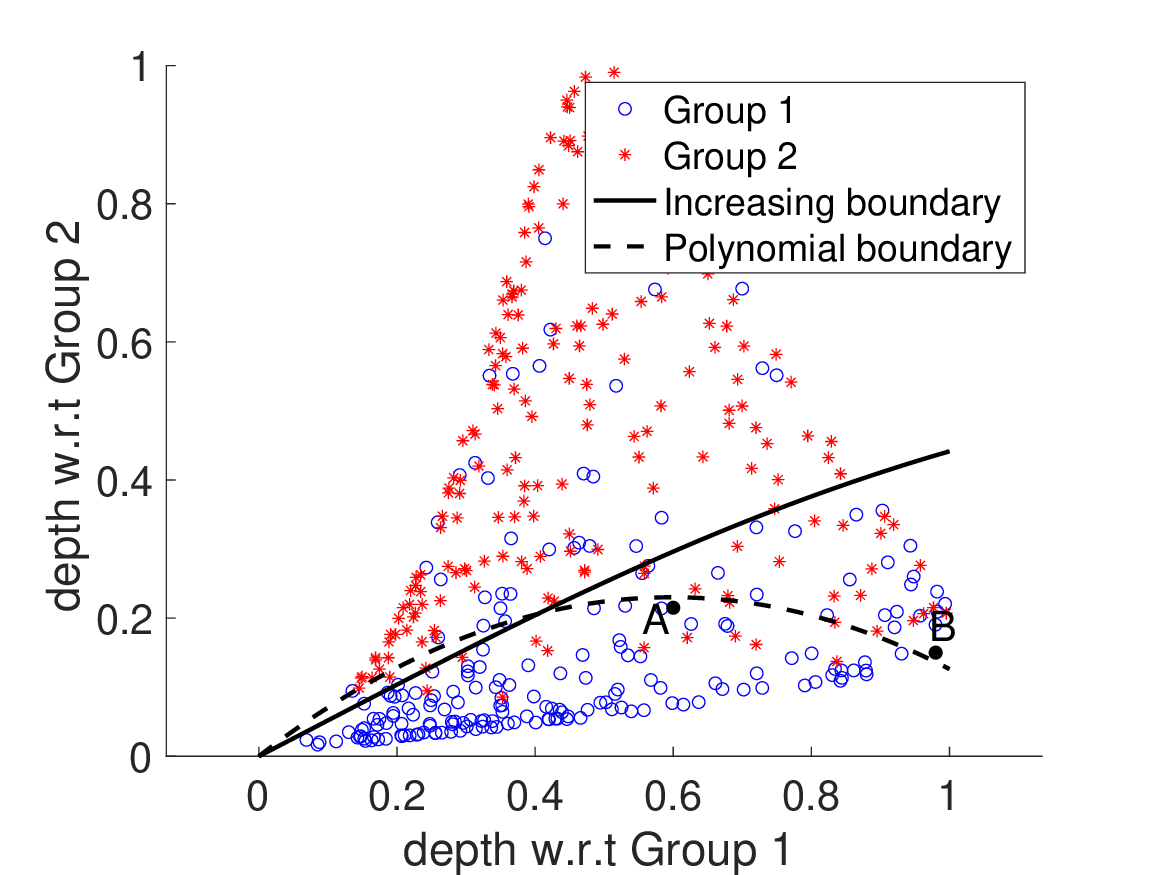}
		\end{minipage}
	}
	\subfigure[misclassification rate]{
		\begin{minipage}[b]{0.47\textwidth}
			\centering
			\includegraphics[scale=0.3]{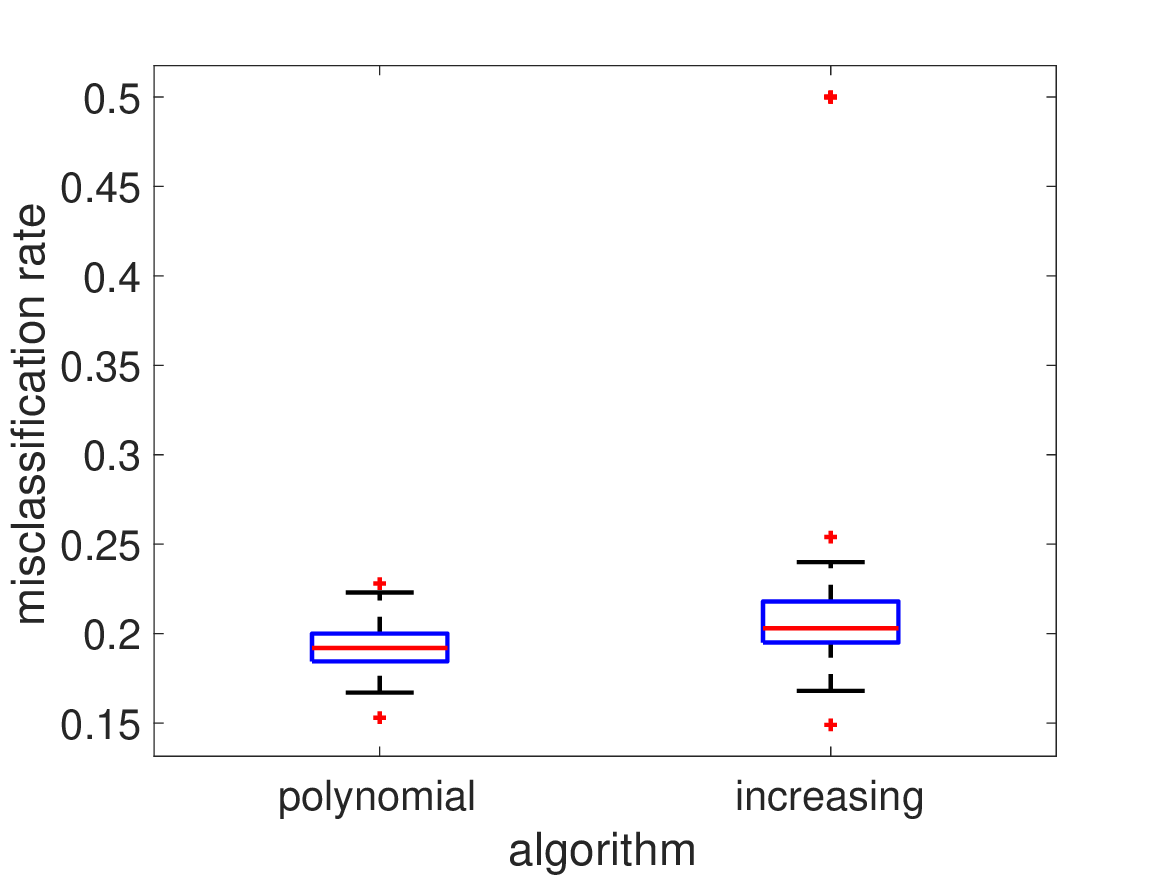}
		\end{minipage}
	}
	\caption{Comparison of the DD classifiers. (a) DD-Classification result.  Data from two groups are represented with blue and red colors, respectivly. The dashed curve represents boundary function obtained by the quadratic function \citep{li2012dd}, and the solid curve represents boundary function obtained from Algorithm \ref{alg:dd_opt}. (b) Boxplots of misclassification rate of 100 experiments from the two methods.}	
	\label{fig:multi_example}
\end{figure*}

Before we move forward to illustrate Algorithm \ref{alg:dd_opt} with spike train simulation examples, we at first compare the classification results of multivariate data with two algorithms, one is the previous algorithm in \citet{li2012dd} with polynomial degree as $2$, where the output will be a quadratic function; the other one is Algorithm \ref{alg:dd_opt} with $k_0=2$, where the output will be a strictly increasing function. We will use the following two groups of data in comparison: 
\begin{eqnarray*}
& &\text{Group 1} \sim N\Bigg(\begin{pmatrix} 0 \\ 0 \end{pmatrix},\begin{pmatrix} 1 & 1 \\ 1 & 4 \end{pmatrix}\Bigg) \\
& &\text{Group 2} \sim N\Bigg(\begin{pmatrix} 1 \\ 1 \end{pmatrix},\begin{pmatrix} 0.25 & 0.25 \\ 0.25 & 1 \end{pmatrix}\Bigg)
\end{eqnarray*}
200 data points for each group will be generated as training set, and 500 data points for each group will be generated as test set to compute the misclassification rate. This experiment is repeated for 100 times and the test misclassification rates are shown in Fig. \ref{fig:multi_example}, where the depth method adopted is the conventional Mahalanobis depth.   

From Fig. \ref{fig:multi_example}(a), one can find that the two boundary functions produce similar classification result. However, the increasing boundary is more interpretable. For example, we consider two data points $\bm{A}$ and $\bm{B}$ shown in Fig. \ref{fig:multi_example}(a). The depth value of $\bm{A}$ is higher than the depth value of $\bm{B}$ with respect to Group 2. In the same time, the depth value of $\bm{A}$ is lower than the depth value of $\bm{B}$ with respect to Group 1. However, the polynomial boundary function assigns $\bm{A}$ to Group 1 and $\bm{B}$ to Group 2, respectively. This classification clearly contradicts the basic notion of depth.  In contrast, the increasing boundary can avoid this problem. Based on Fig. \ref{fig:multi_example}(b), the overall misclassification rates of the two algorithms are close to each other. The medians of misclassification rates of polynomial and increasing boundary are $19.1\%$ and $20.2\%$ respectively. 
Finally, the computation efficiency of two algorithms can be summarized in Table \ref{tab:com_time}.  We can see that Algorithm \ref{alg:dd_opt} is slightly more efficient. 
\begin{table}[h!]
\centering
\begin{tabular}{|c|c|c|c|}
\hline
\textbf{Boundary function} & Mean & Standard deviation  \\ \hline
polynomial & 1.65 & 4.22  \\ \hline
increasing & 0.94 & 1.29  \\ \hline

\end{tabular}
\caption{Computational efficiency of the two algorithms (in units of seconds). }
\label{tab:com_time}
\end{table}

In conclusion, the DD classifier with increasing function in Algorithm \ref{alg:dd_opt} is more interpretable and time efficient, whereas it shares roughly the same performance with the algorithm in \citet{li2012dd}. 

\section{Experimental Results} \label{sec:results}
In this section, we will use two simulation examples and two real experimental datasets to illustrate the proposed methods on spike train data.  In particular, we will systematically compare our new classification with commonly used classifiers.  All the methods are listed as follows: 
\begin{itemize}
\item DD: DD classifier with Algorithm \ref{alg:dd_opt}. 
\item MD: Maximum Depth classifier \citep{liu1990notion}, where classification is simply based on maximal depth value. 
\item LM: classical Likelihood Method, where the classification is based on Gaussian likelihood of firing rate vector in discretized time bins \citep{hatsopoulos2001representations}.    
\item MM1: Minimum distance to the Mean spike train \citep{wu2013estimating}. 
\item MM2: Minimum distance to the Median spike train. Similar to MM1 except for using newly developed median spike train. 
\end{itemize}

\begin{figure*} [h!]
	\centering
	\subfigure[HPP]{
		\begin{minipage}[b]{0.47\textwidth}
			\centering
			\includegraphics[scale=0.3]{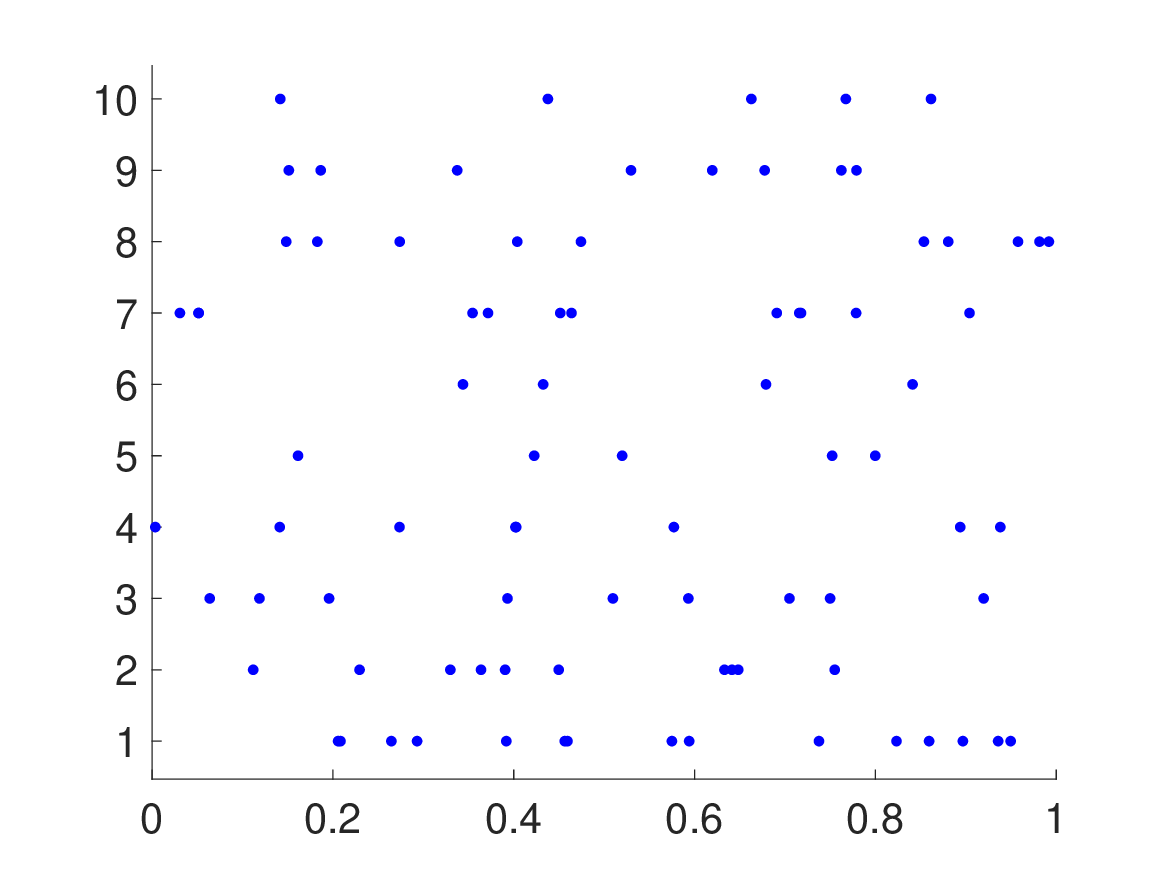}
		\end{minipage}
	}
	\subfigure[IPP]{
		\begin{minipage}[b]{0.47\textwidth}
			\centering
			\includegraphics[scale=0.3]{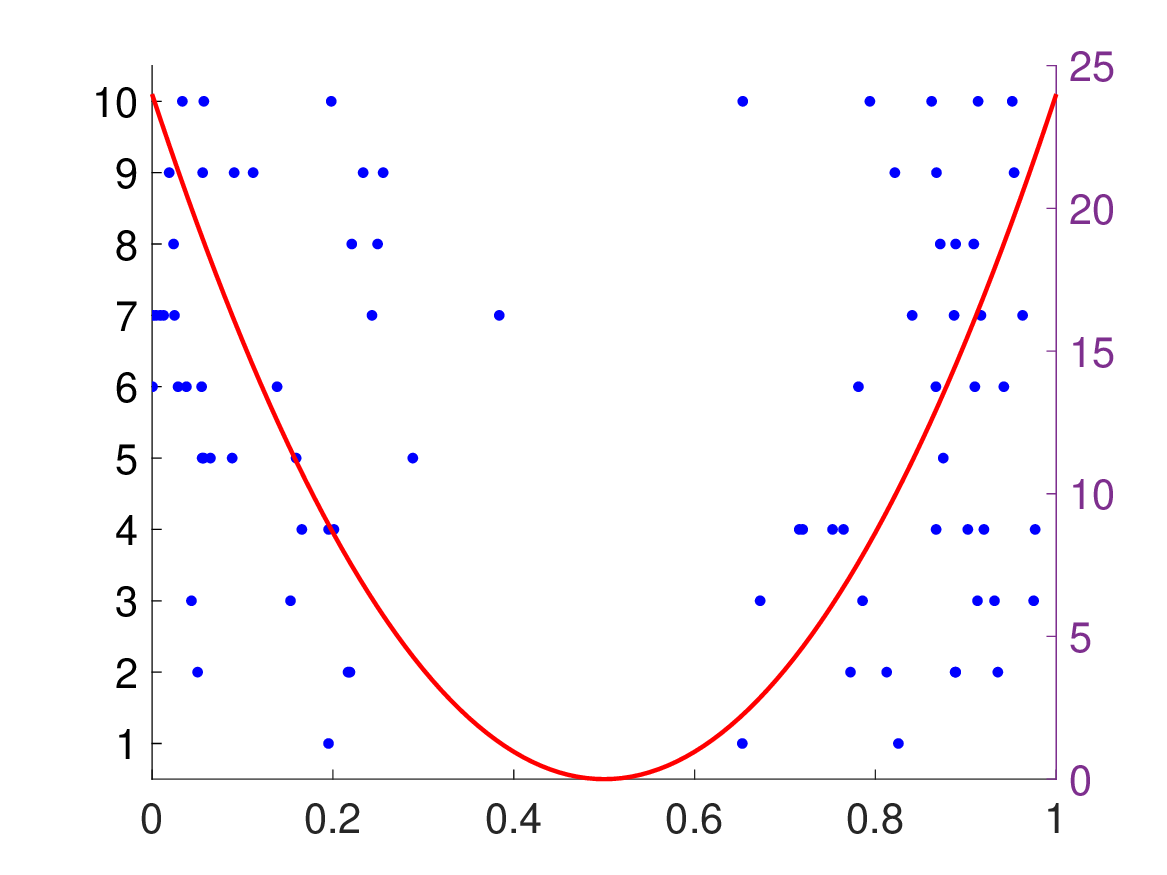}
		\end{minipage}
	}
	\\
	\subfigure[median and mean in Group I]{
		\begin{minipage}[b]{0.47\textwidth}
			\centering
			\includegraphics[scale=0.3]{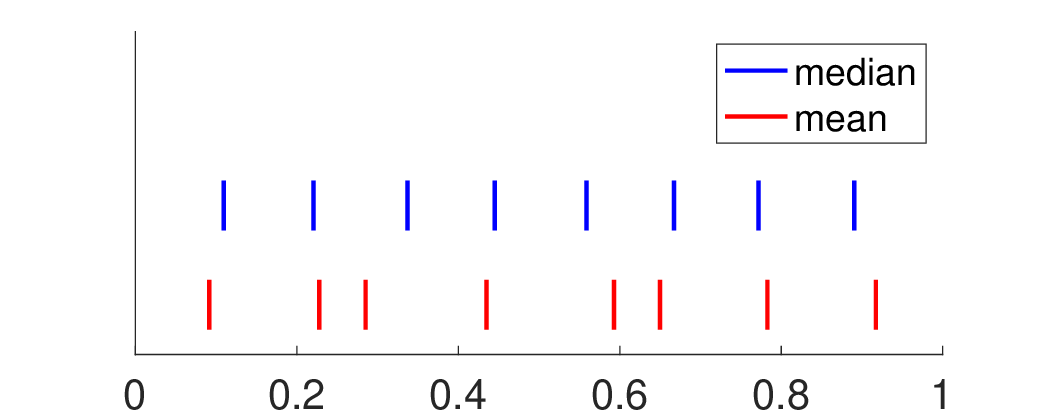}
		\end{minipage}
	}
	\subfigure[median and mean in Group II]{
		\begin{minipage}[b]{0.47\textwidth}
			\centering
			\includegraphics[scale=0.3]{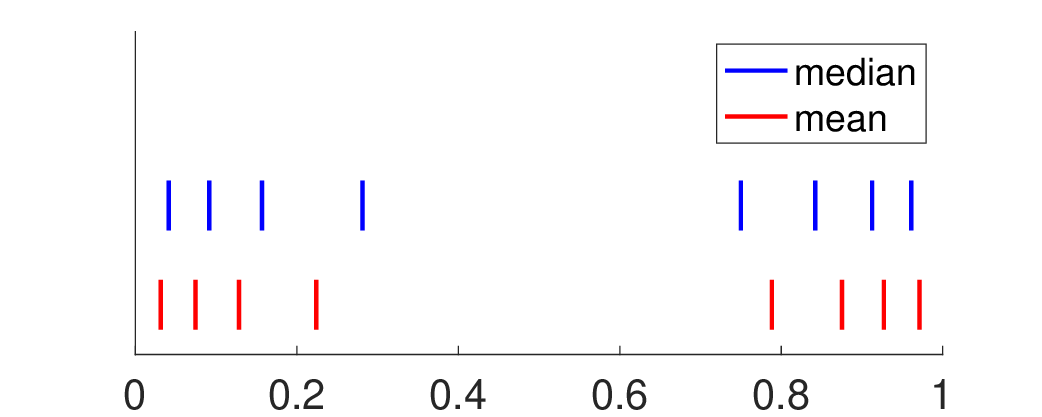}
		\end{minipage}
	}
	\caption{Realizations of spike trains from the two groups and the median and mean estimations.  (a) 10 realization from Group I. Each row is one realization and the blue dot denotes a spike event. (b) 10 realization from Group II. Each row is one realization and the blue dot denotes a spike event. The indices of spike trains are shown in the left y-axis. The red curve is the intensity function, and its value is shown along the right y-axis. (c) The mean and median spike trains in Group I with sample size 500. Each vertical line represents one spike.  The blue lines are the spike events in the median, and the red lines are the spike events in the mean. (d) Same as (c) except for Group II.}	
	\label{fig:HPPvsIPP}
\end{figure*}

\subsection{Simulation Example 1: HPP vs. IPP}
We consider the following two groups of spike trains on the time domain $[0, 1]$: 
\begin{itemize}
\item Group I: HPP with constant intensity rate $\lambda=8$. 
\item Group II: IPP with intensity function $\lambda(t)=96\big(t-\frac{1}{2}\big)^2$. 
\end{itemize}
It is easy to verify that the total intensity in each group (I or II) is 8, which is the mean number of events in a realization.  We simulate 500 realizations in each group. 10 example realizations for each group are shown in Fig. \ref{fig:HPPvsIPP}(a) and (b), respectively. One can find the HPPs are located uniformly within the time domain, while the IPPs are mainly clustered at the two boundary sides.

The mean and median of Group I and II are shown in Fig. \ref{fig:HPPvsIPP}(c) and (d), respectively. The cardinalities for the median and mean are 8 in both groups. In additional, for Group I, median is more uniformly located than mean, which indicates that median shows a better description of homogeneous pattern in the spike train data. For Group II, both median and mean have reasonable locations and can capture the parabolic shape of the intensity function.  

\begin{figure*} [h!]
	\centering
	
	\subfigure[DD classifier in training data]{
		\begin{minipage}[b]{0.47\textwidth}
			\centering
			\includegraphics[scale=0.3]{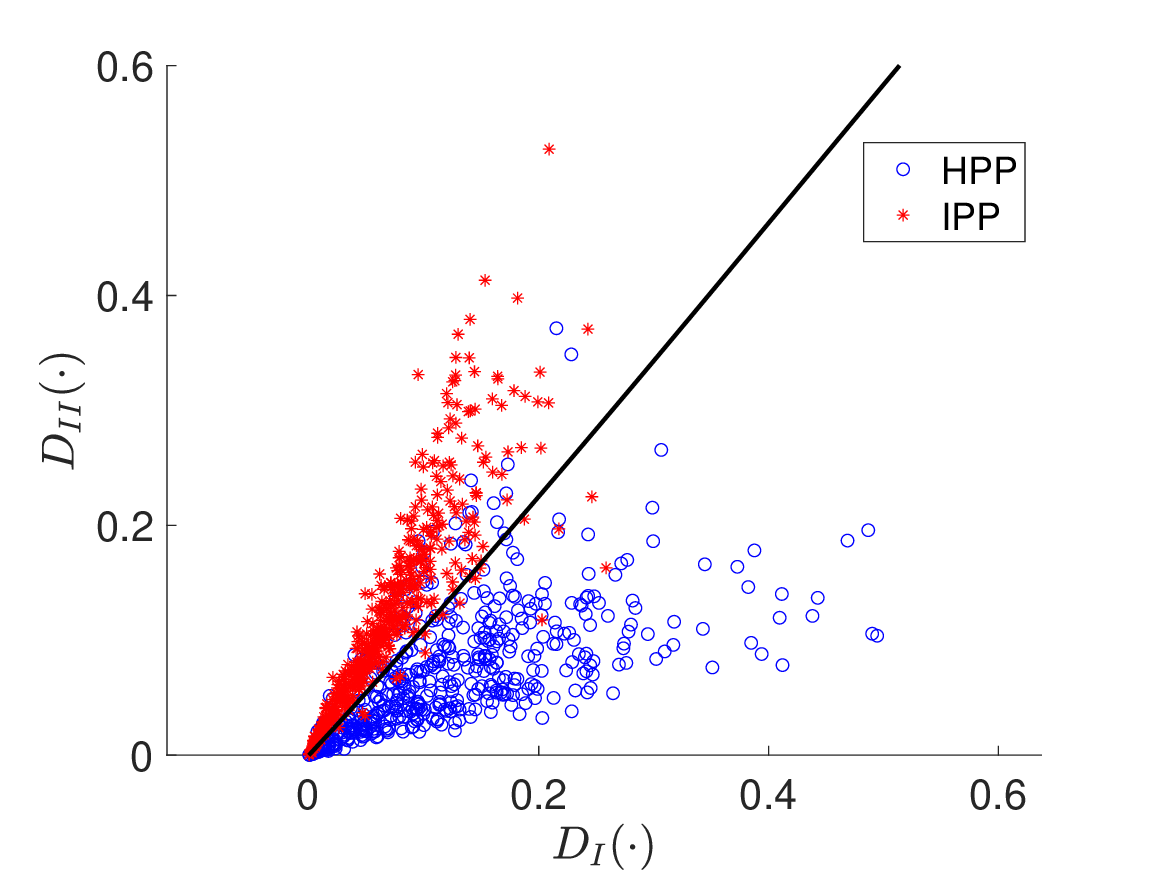}
		\end{minipage}
	}
	\subfigure[DD classifier in test data]{
		\begin{minipage}[b]{0.47\textwidth}
			\centering
			\includegraphics[scale=0.3]{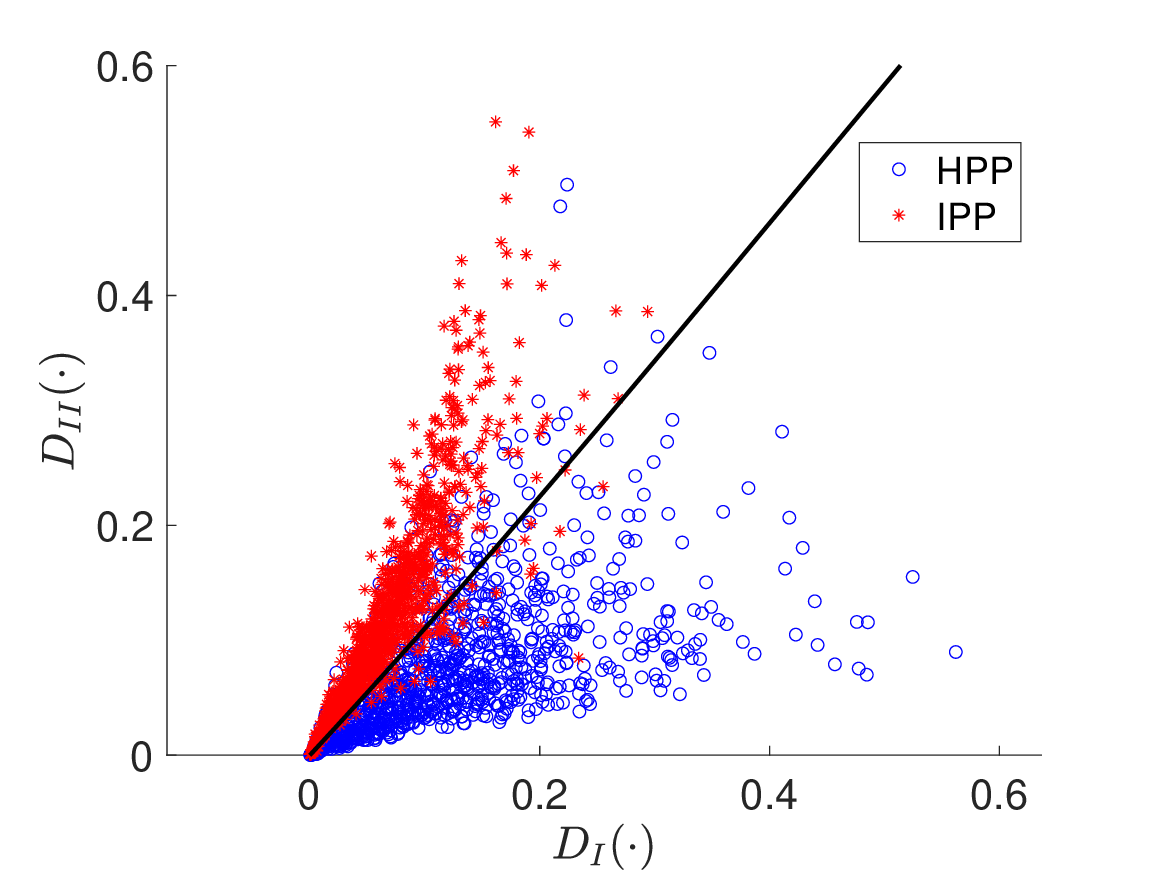}
		\end{minipage}
	}
	
	\subfigure[boxplots of misclassification rate]{
    		\begin{minipage}[b]{0.47\textwidth}
			\centering
   		 	\includegraphics[scale=0.3]{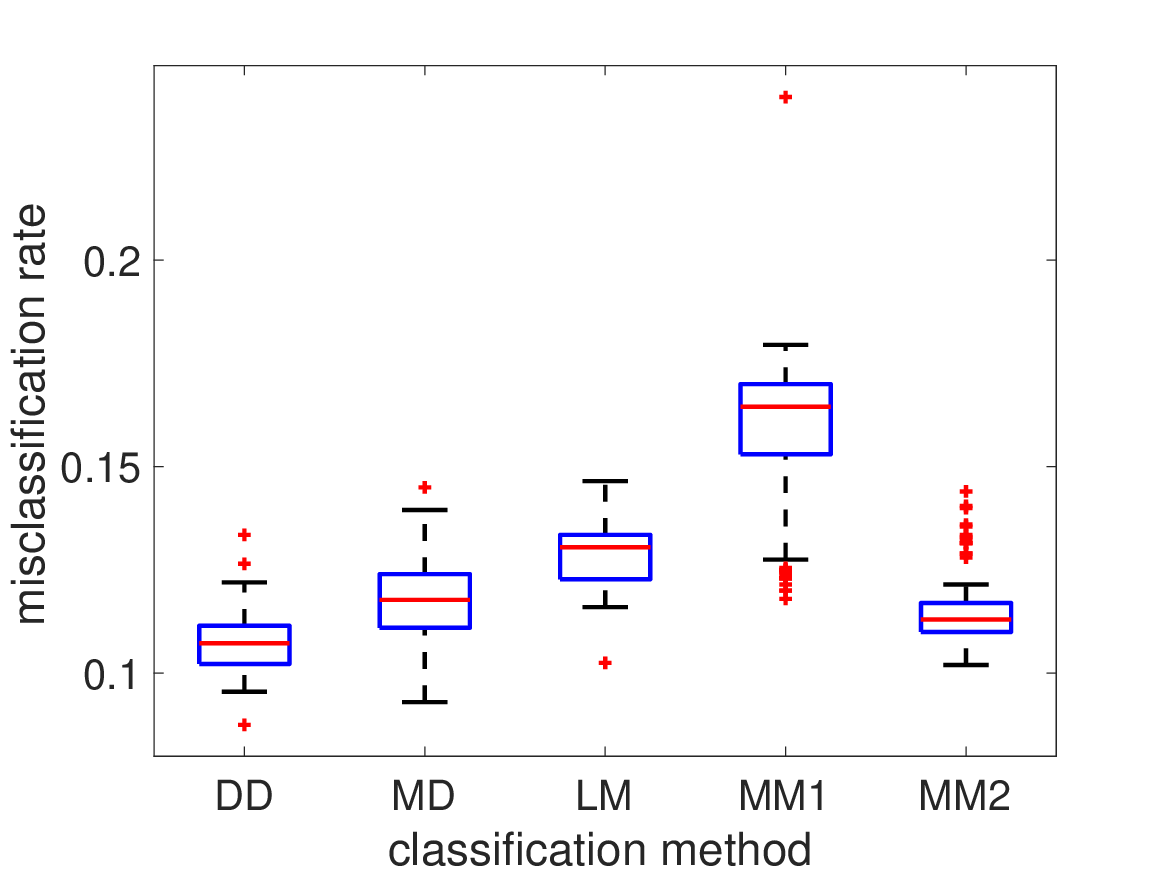}
    		\end{minipage}
    	}
	\subfigure[example outliers in each group]{
    		\begin{minipage}[b]{0.47\textwidth}
			\centering
   		 	\includegraphics[scale=0.3]{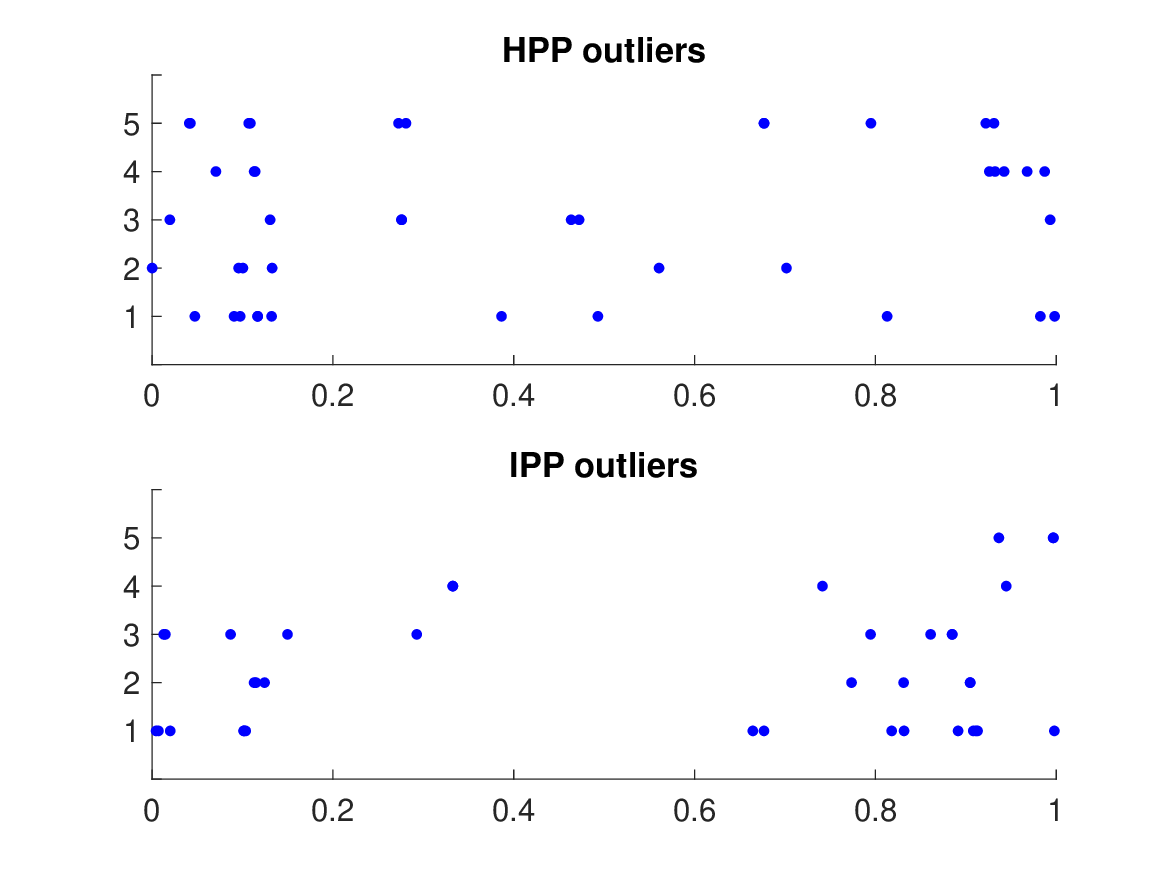}
    		\end{minipage}
    	}
	\caption{Result of spike train classification of HPP vs IPP simulations. (a) Boundary function of DD classifier and DD plot for one experiment of training data, the black curve is the optimal boundary function. (b) Same as (a) except for the test data. (c) Boxplots of misclassification rate of test data for different classifiers. (d) 5 detected outlier examples in each group. Each row is one spike train and each blue dot represents a spike event. }
	\label{fig:classify_sim}
\end{figure*}


To evaluate classification performance on these two groups, we will use the simulated  500 spike trains in each group as training data, and then simulate 1000 new spike trains in each group as test data. The hyper-parameter $k_0$ in Algorithm \ref{alg:dd_opt} is set to be $5$, the initial guess of $\bm{a}$ is a column with all entries $0$, and the hyper-parameter $r$ in Definition \ref{def:wholedef} is 1. The classification result is shown in Fig. \ref{fig:classify_sim}.  We repeat the experiment 100 times to evaluate the distribution of the misclassification errors using boxplots in Fig. \ref{fig:classify_sim}(c). 

From Fig. \ref{fig:classify_sim}(a) and \ref{fig:classify_sim}(b), one can conclude that the two groups of spike train, HPP and IPP, can be effectively distinguished by the DD classifier.  Fig. \ref{fig:classify_sim}(c) shows the result of misclassification rate with respect to the DD classifiers, as well as with other commonly used classifiers. It is clear that the DD classifier outperforms all the others. In particular, the median of misclassification rate of the DD classifier is 0.1072, lower than the medians of other methods.  

We also adopt the outlier detection method to identify outliers in each training group by setting the threshold $\delta=0.01$ in Section \ref{sec:out_det}, then we remove the outliers and retrain the classifiers. This outlier removal step is aimed to delete the unusual spike trains in each training group, thereby improve the training model and increase the classifier accuracy. Some examples of detected outliers are shown in Fig. \ref{fig:classify_sim}(d). Most of them are due to extreme-adjacency of two spikes. We then conduct classification by removing those outliers in the training data and the same classification procedure in the original case is applied to the test data.  We find that the median of the misclassification rate of the DD classifier is 0.1015, which is slightly better than the rate 0.1072 without outlier deletion. For each experiment, the amount of detected outlier is small compared with the group size (often less than $1.5\%$ of the group size). The classification result will not be significantly improved by outlier deletion. Thus, this $0.57\%$ improvement shows the effectiveness of our outlier detection framework.

\subsection{Simulation example 2: IPP vs. Hawkes process}
We consider the following two groups of spike train on the time domain $[0, 1]$: 
\begin{itemize}
\item Group I: IPP with intensity function $\lambda(t)=\frac{100}{\sqrt{2\pi}}e^{-\frac{(t-0.25)^2}{2\times 0.05^2}}I(t\leq 0.5)+\frac{100}{\sqrt{2\pi}}e^{-\frac{(t-0.75)^2}{2\times 0.05^2}}I(t>0.5)$, where $I(\cdot)$ is the indicator function. 
\item Group II: Hawkes process  with conditional intensity function $\lambda^*(t\arrowvert H_t)=\frac{1}{2}\lambda(t)+\sum_{t_i<t}\alpha e^{-\beta(t-t_i)}$, where $\lambda(t)$ is the intensity function in Group I, $\alpha=15$, $\beta=30$, $t_i$ is the time events before time $t$, and $H_t$ represents the history of time events up to time $t$. 
\end{itemize}

\begin{figure*} [h!]
	\centering
	\subfigure[IPP]{
		\begin{minipage}[b]{0.47\textwidth}
			\centering
			\includegraphics[scale=0.3]{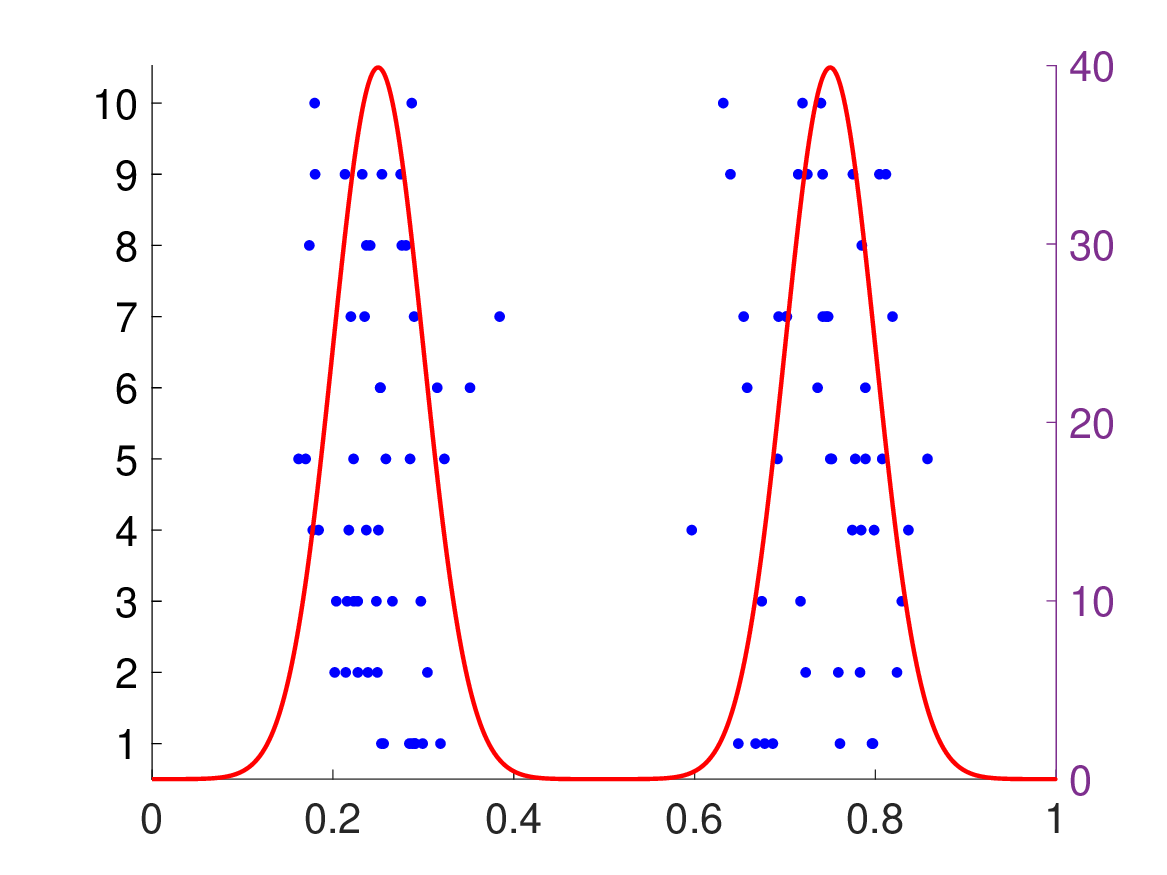}
		\end{minipage}
	}
	\subfigure[Hawkes process]{
		\begin{minipage}[b]{0.47\textwidth}
			\centering
			\includegraphics[scale=0.3]{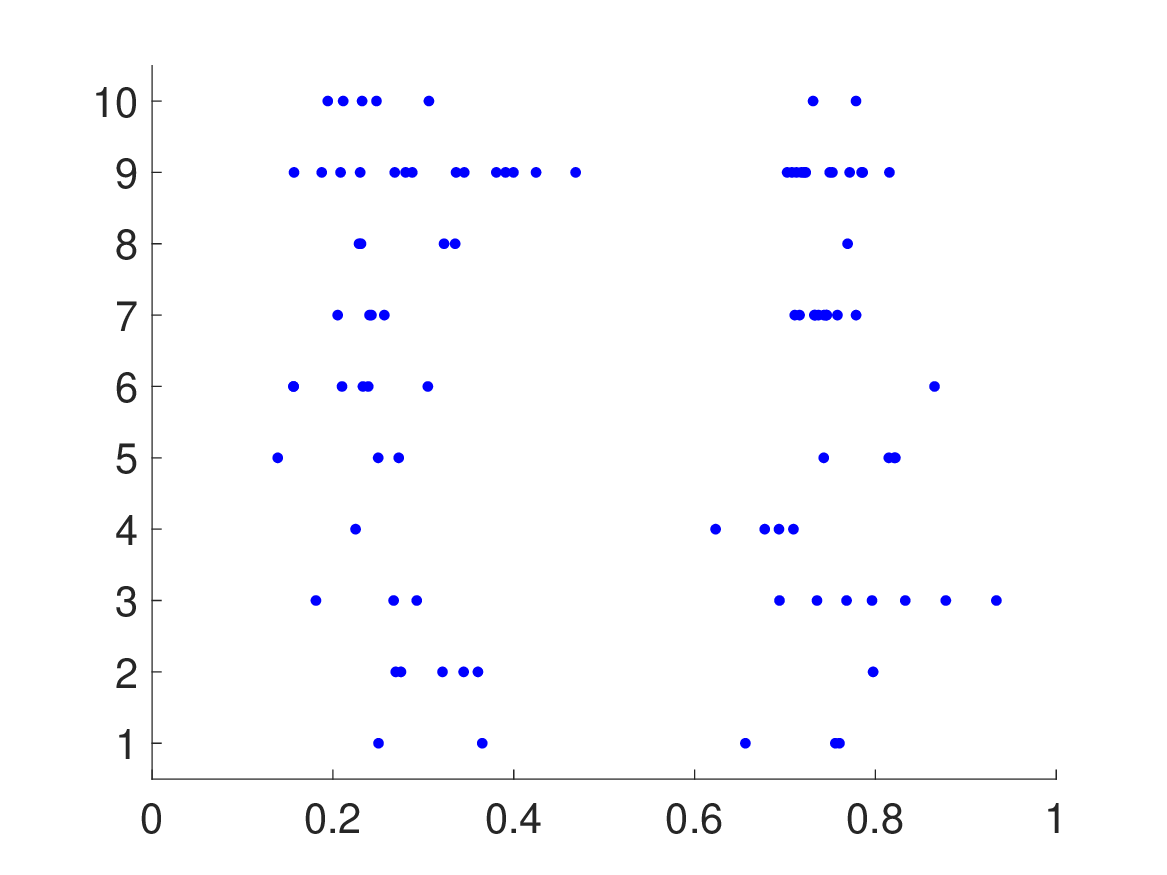}
		\end{minipage}
	}
	\\
	\subfigure[median and mean in Group I]{
		\begin{minipage}[b]{0.47\textwidth}
			\centering
			\includegraphics[scale=0.3]{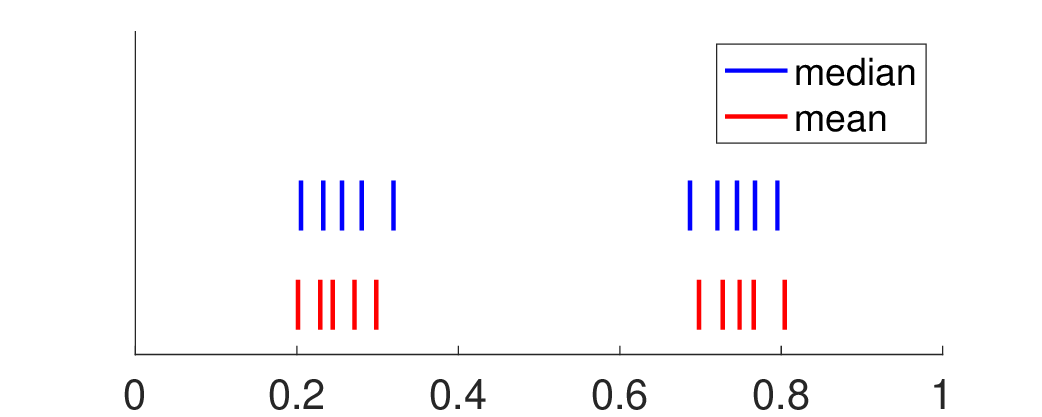}
		\end{minipage}
	}
	\subfigure[median and mean in Group II]{
		\begin{minipage}[b]{0.47\textwidth}
			\centering
			\includegraphics[scale=0.3]{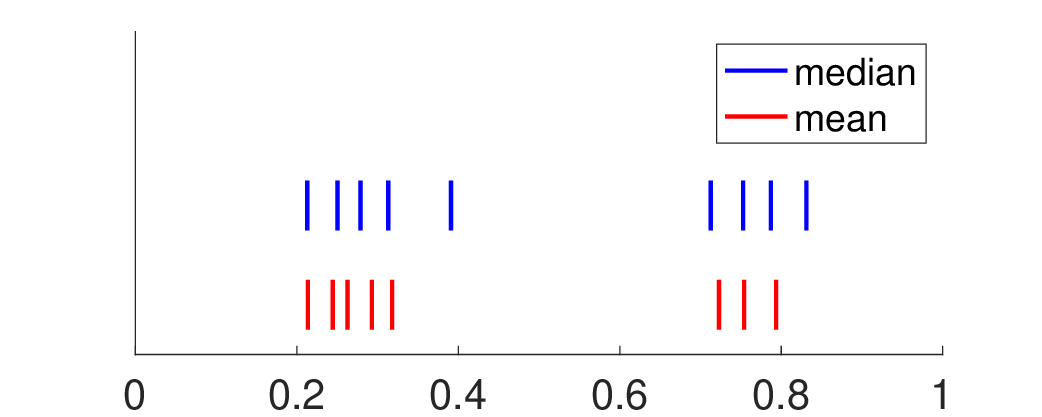}
		\end{minipage}
	}
	\caption{Realizations of spike trains from the two groups and the median and mean estimations.  (a) 10 realizations from Group I. Each row is one realization and the blue dot denotes a spike event, where the left y-axis shows the indices of spike trains. The red curve is the intensity function and its value is shown along the right y-axis. (b) 10 realizations from Group II. Each row is one realization and the blue dot denotes a spike event. (c) The mean and median spike trains in Group I with sample size 500. Each vertical line represents one spike.  The blue lines are the spike events in the median, and the red lines are the spike events in the mean. (d) Same as (c) except for Group II.}	
	\label{fig:IPP_hawkes}
\end{figure*}

We simulate 500 independent realizations in each group.  10 example realizations in both groups are shown in Fig. \ref{fig:IPP_hawkes}(a) and (b), respectively. According to the simulation process, spike trains from Group I have more spikes near 0.25 and 0.75, respectively.  
In contrast, the spikes in Group II are still around these two time regions, ableit with more variability due to the history dependency. In addition, we find the mean cardinality of two groups are close to each other, which increases the difficulty of classification procedures.

The median and mean spike trains in both groups are shown in Fig. \ref{fig:IPP_hawkes}(c) and (d), respectively.  We can clearly see the clustering property in these summary statistics on templates. Therefore, it is not straightforward to distinguish two groups by the number of spikes or spike time locations. However, we will show that the DD classifier is able to provide a reasonable performance. 

Same as the previous section, we will use the 500 spike trains in each group as training data, and then simulate 1000 new spike trains in each group as test data. When applying Algorithm \ref{alg:dd_opt}, the hyper-parameter $k_0$ is $5$ and the initial value of $\bm{a}$ is a column with all entries $0$.  The hyper-parameter $r$ in Definition \ref{def:wholedef} is 1. The experiments is repeated 100 times and we use boxplot to summarize the misclassification error rates with different classifiers.  We will use the same classifiers as before except adding one in this simulation example: 
\begin{itemize}
\item IA: Treat the Hawkes process as IPP and estimate its intensity function using kernel methods, and then conduct the DD classifier with Algorithm \ref{alg:dd_opt}. 
\end{itemize}

\begin{figure*} [h!]
	\centering
	
	\subfigure[training result]{
		\begin{minipage}[b]{0.47\textwidth}
			\centering
			\includegraphics[scale=0.3]{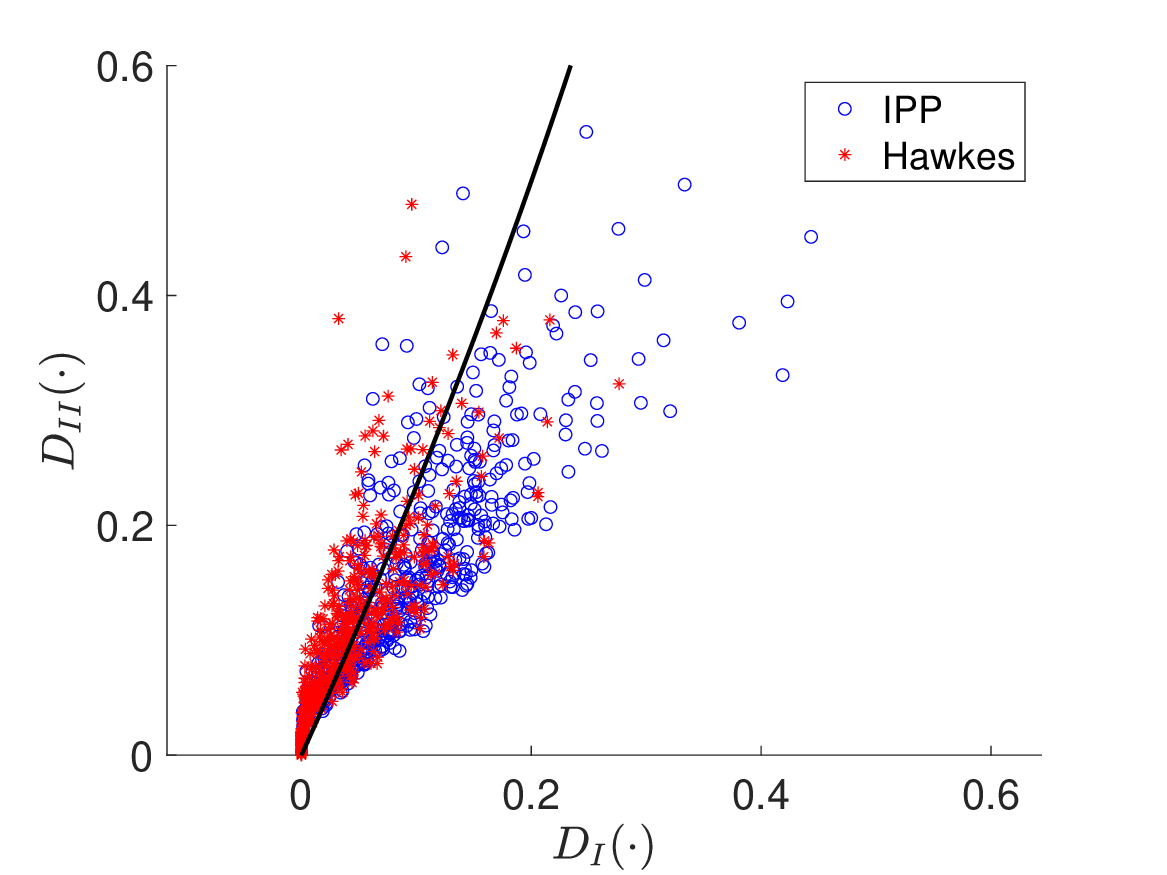}
		\end{minipage}
	}
	\subfigure[test result]{
		\begin{minipage}[b]{0.47\textwidth}
			\centering
			\includegraphics[scale=0.3]{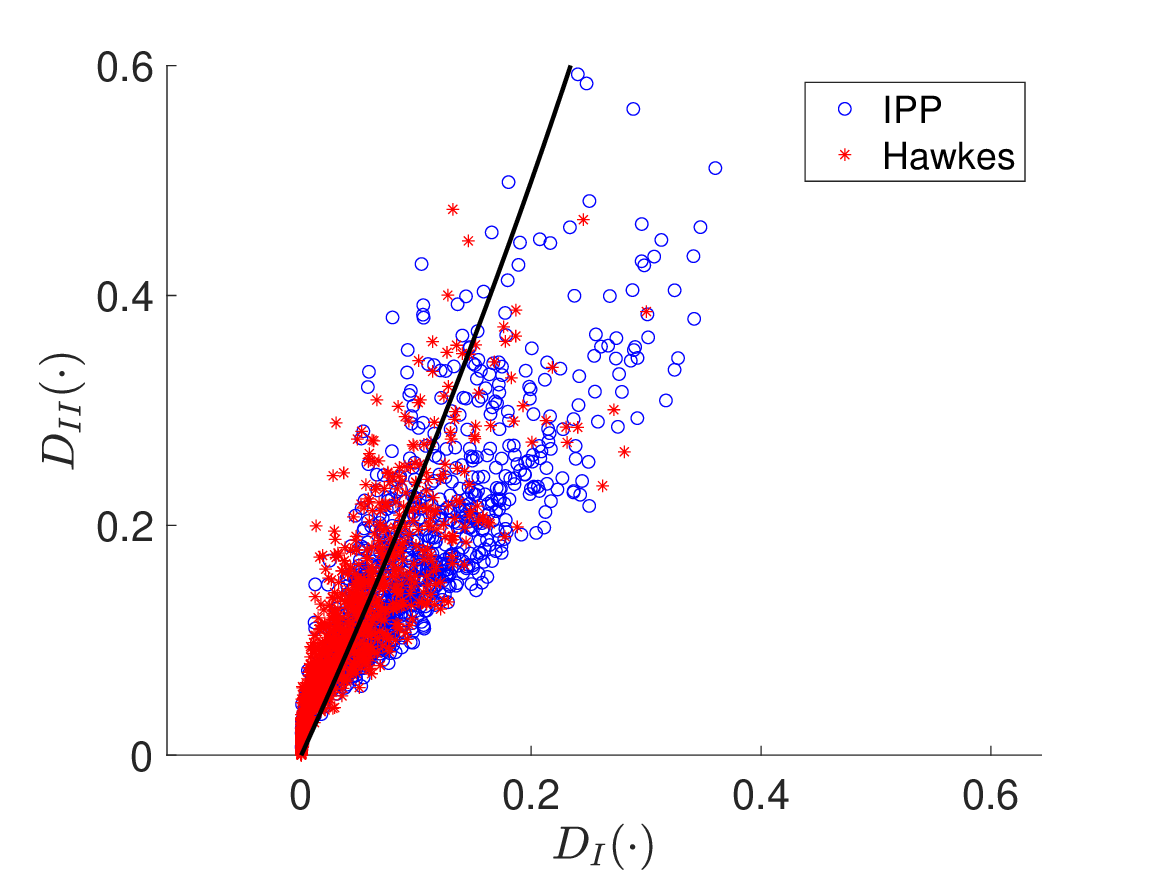}
		\end{minipage}
	}
	\\
	\subfigure[boxplots of misclassification rate]{
    		\begin{minipage}[b]{0.47\textwidth}
			\centering
   		 	\includegraphics[scale=0.3]{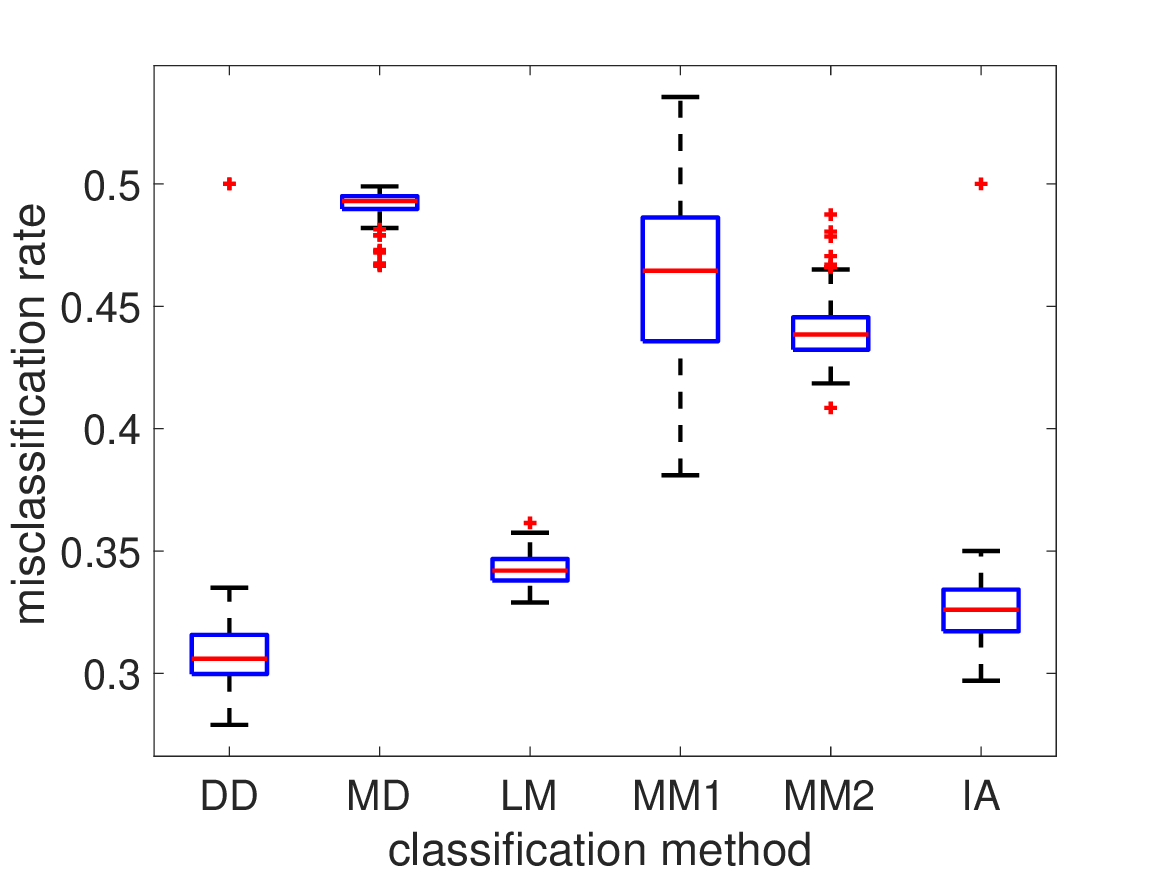}
    		\end{minipage}
    	}
	\subfigure[examples of detected outliers]{
    		\begin{minipage}[b]{0.47\textwidth}
			\centering
   		 	\includegraphics[scale=0.3]{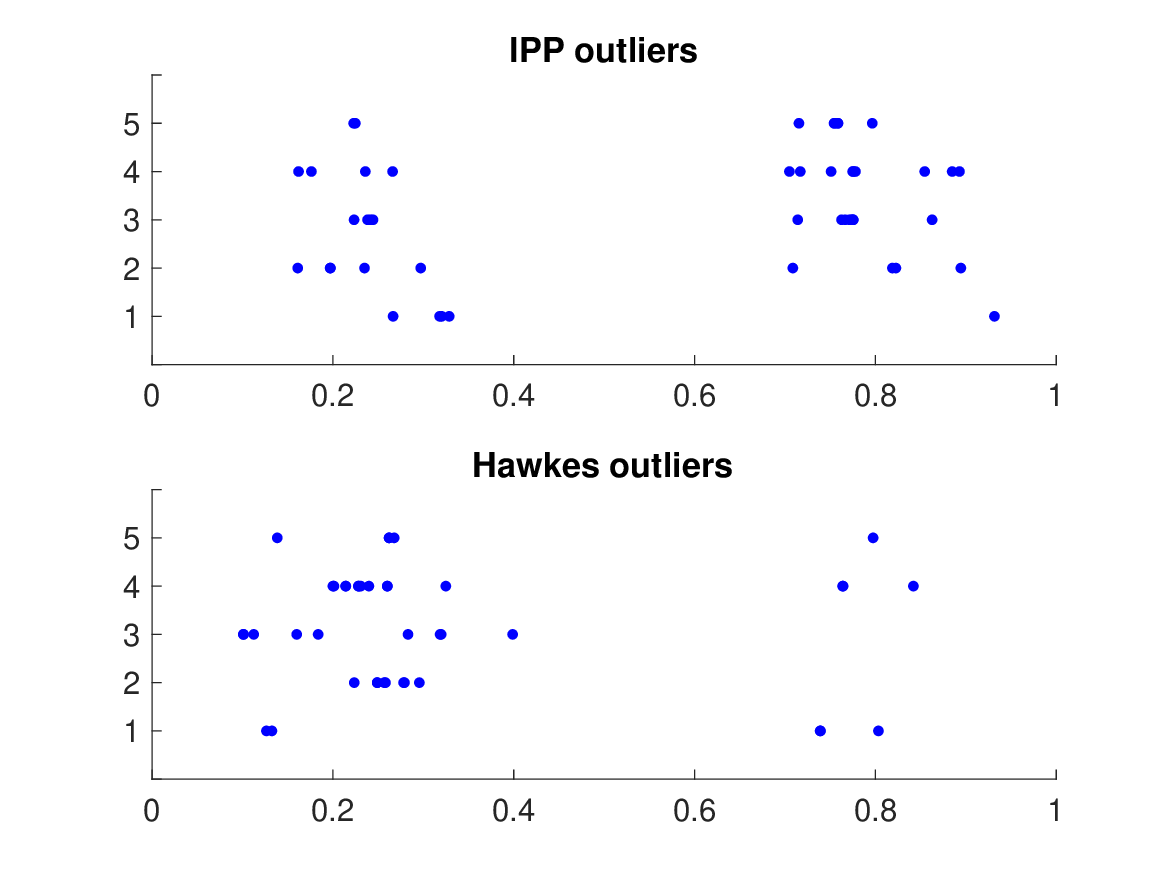}
    		\end{minipage}
    	}
	\caption{Result of spike train classification of IPP vs Hawkes process. (a) DD plot in the training data in the two groups, where blue and red circles are spike trains from IPP and Hawkes process, respectively, and the black curve is the optimal boundary function. (b) Same as (a) except for the same experiment of test data. (c) Boxplots of misclassification rate of test data for six classifiers.  (d) Five example detected outliers for either group, where each blue dot represents a spike event.}
	\label{fig:classify_IPP_hawkes}
\end{figure*}

All results are shown in Fig. \ref{fig:classify_IPP_hawkes}. 
From Fig. \ref{fig:classify_IPP_hawkes}(a) and (b), it can be found that there is a strong overlap of depth values of the Poisson process and Hawkes process.  The depth values of the Hawkes process remain slightly lower in most trials. The DD classifier can capture the distinction between two groups and return an appropriate boundary function. In Fig. \ref{fig:classify_IPP_hawkes}(c), the median of misclassification rate of the DD classifier is 0.3060, which shows a more successful result compared with other classifiers.  Since most points in DD plot are located at the up-left part, the classification result is very poor for the MD classifier. The result of IA is slightly worse than that of DD, which shows the feasibility of Poisson process approximation. Both MM1 and MM2 show poor performance and their misclassification errors on some experiments are higher than 0.5, the random guess error rate. That is, the metric approach may not be stable for spike train analysis. Finally, same as in the previous simulation, we adopt one more step, outlier deletion with $\delta=0.01$, for training data before classification. Some outlier samples are shown in Fig. \ref{fig:classify_IPP_hawkes}(d). Consistent to the result in previous section, the median of misclassification rate of DD slightly decreases to 0.3008. This again supports the usefulness of our outlier detection framework.

\subsection{Real dataset 1: macaque prefrontal cortex}
We will now use one real experimental dataset to demonstrate the proposed depth-based methods. This is for neuronal spike train recordings from macaque prefrontal cortex during a somatosensory working memory task. This study was originally conducted in \citet{romo1999neuronal,brody2003timing}, and the dataset was taken from \citet{romo2016single}. In this experiment, the monkey was stimulated by electrodes with different frequencies and the spike times of 7 cells were recorded. To conduct our framework, we first choose a certain cell. Then, we collect the spikes 500 seconds after the 10 Hz and 34 Hz frequencies stimuli were released. The spikes after 10 Hz are the slightly stimulated spikes for Group I; while the spikes after 34 Hz are the strongly stimulated spikes for Group II. 
\begin{figure*} [h!]
	\centering
	\subfigure[Group I]{
		\begin{minipage}[b]{0.47\textwidth}
			\centering
			\includegraphics[scale=0.3]{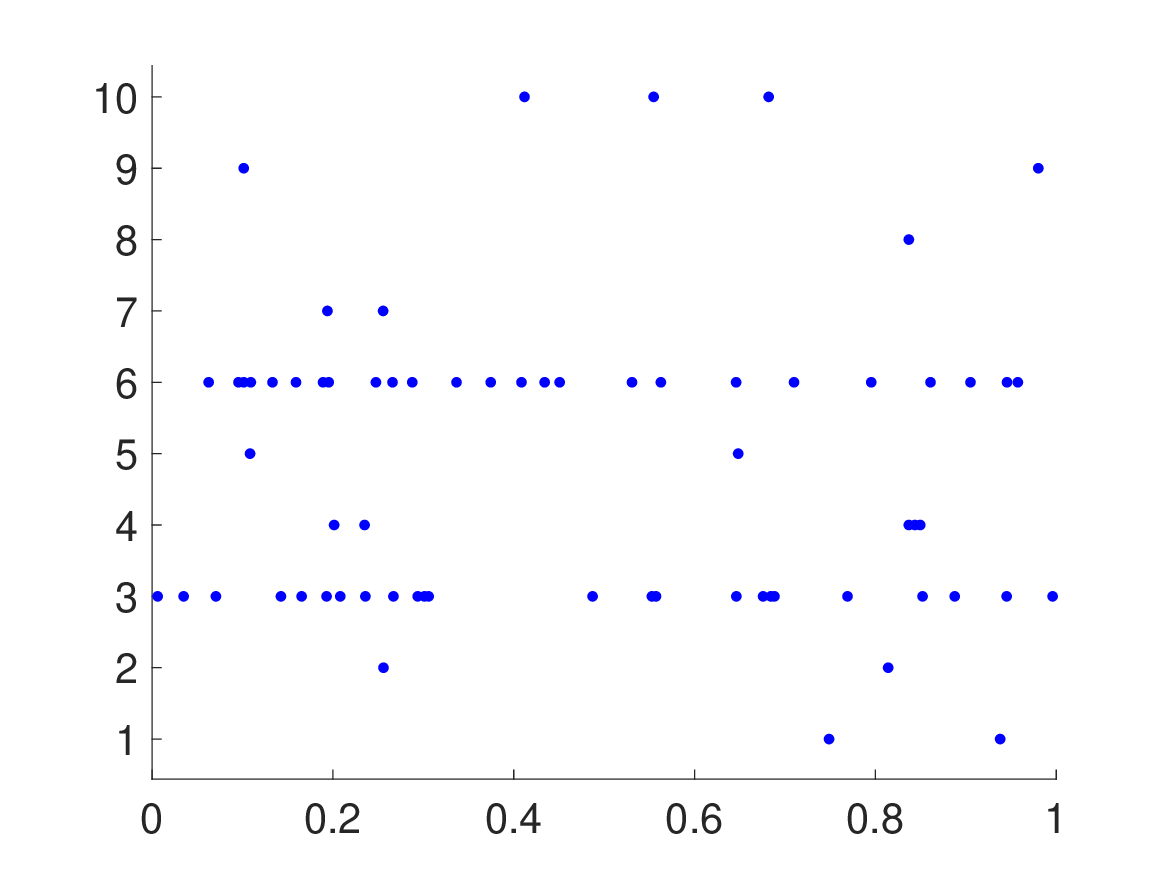}
		\end{minipage}
	}
	\subfigure[Group II]{
		\begin{minipage}[b]{0.47\textwidth}
			\centering
			\includegraphics[scale=0.3]{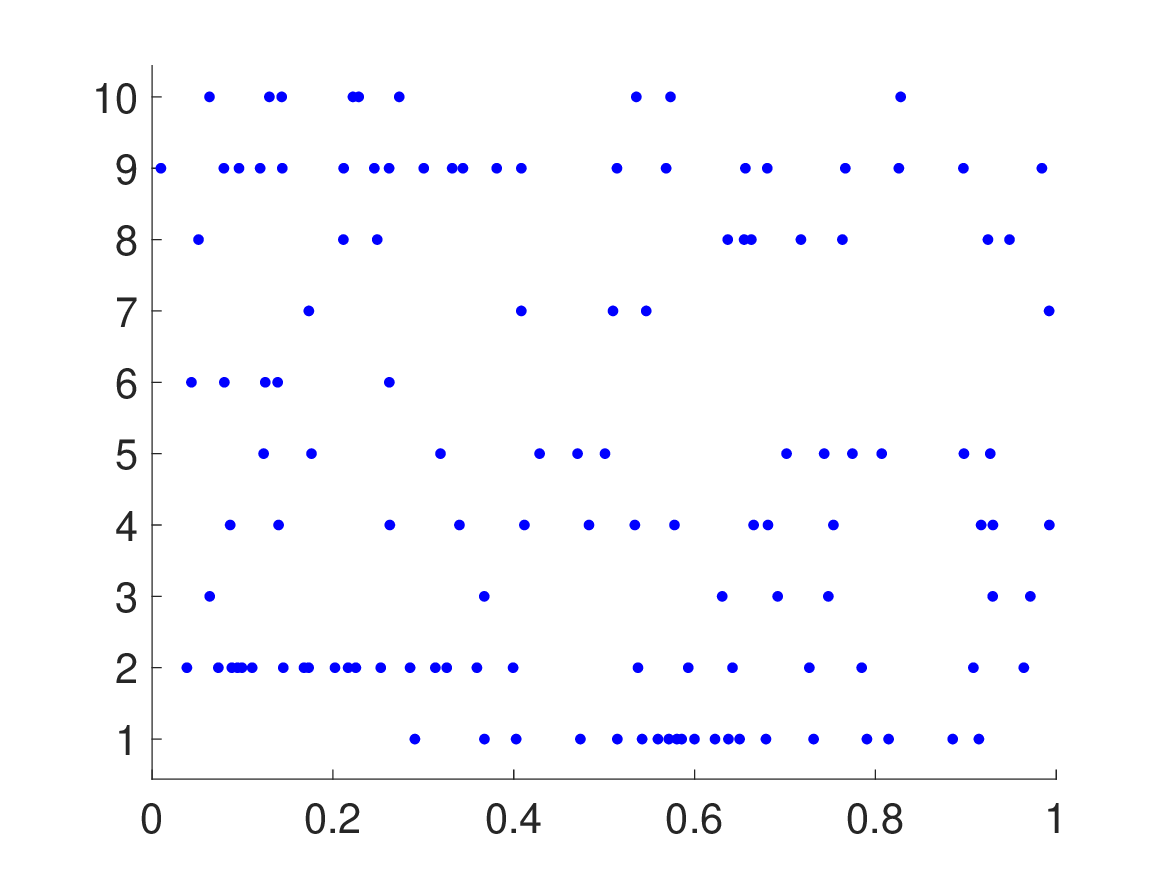}
		\end{minipage}
	}
	\caption{10 example trials in training data. (a) 10 randomly selected training trials from Group I. Each row represents a trial and each blue point represents a spike event, the y-axis shows the indices of trials. (b) Same as (a) except for Group II. }	
	\label{fig:real_reali}
\end{figure*}
The sample size is 399 in Group I and 397 in Group II. To simplify the analysis, we normalize the time domain of each group as $[0,1]$. Next, for each group, we randomly select 297 trials as training data and the remaining trials as test data. Thus, the ratio of training size and test size is approximately $3:1$. Some example trials in the training data are shown in Fig. \ref{fig:real_reali}. 
We note that most spike trains in Group I have smaller cardinalities compared with spike trains in Group II. 
However, there are still spike trains in Group I having large cardinalities, which may increase difficulty in the classification task. 
\begin{figure*} [h!]
	\centering
	
	\subfigure[mean and median of Group I]{
		\begin{minipage}[b]{0.47\textwidth}
			\centering
			\includegraphics[scale=0.3]{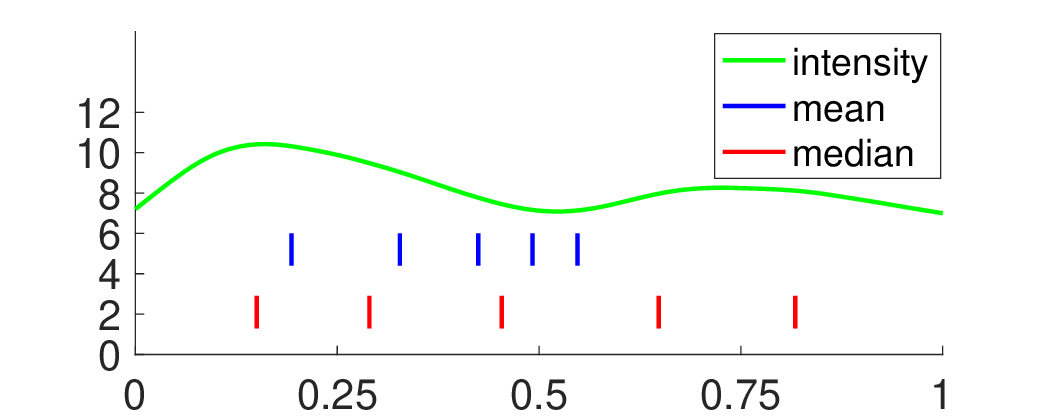}
		\end{minipage}
	}
	\subfigure[mean and median of Group II]{
		\begin{minipage}[b]{0.47\textwidth}
			\centering
			\includegraphics[scale=0.3]{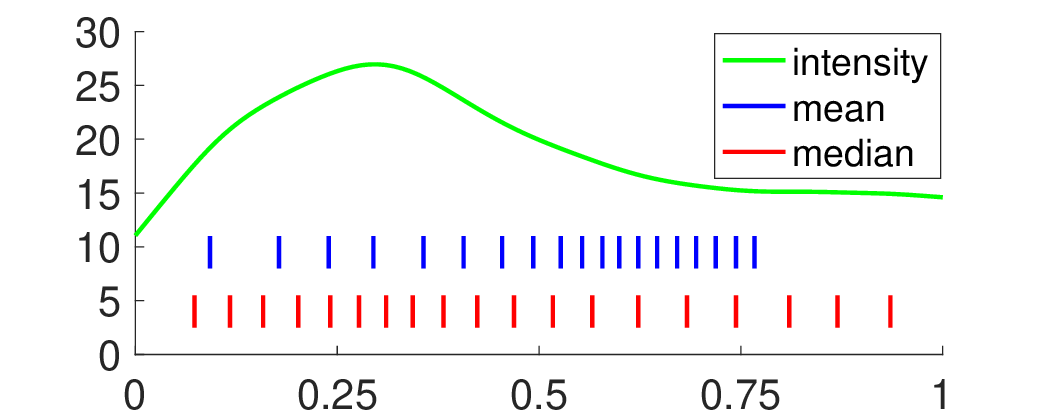}
		\end{minipage}
	}
	\\
	\subfigure[training result]{
    		\begin{minipage}[b]{0.47\textwidth}
			\centering
   		 	\includegraphics[scale=0.3]{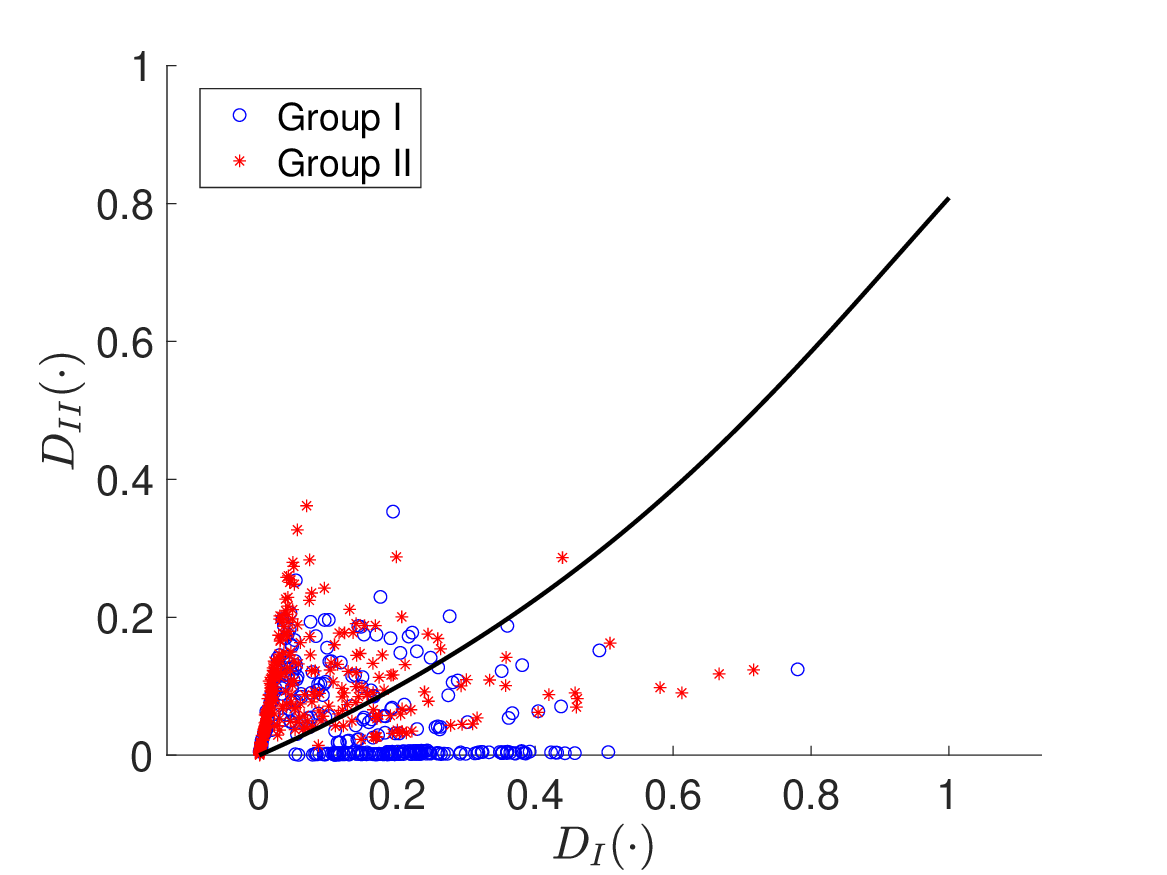}
    		\end{minipage}
    	}
	\subfigure[test result]{
    		\begin{minipage}[b]{0.47\textwidth}
			\centering
   		 	\includegraphics[scale=0.3]{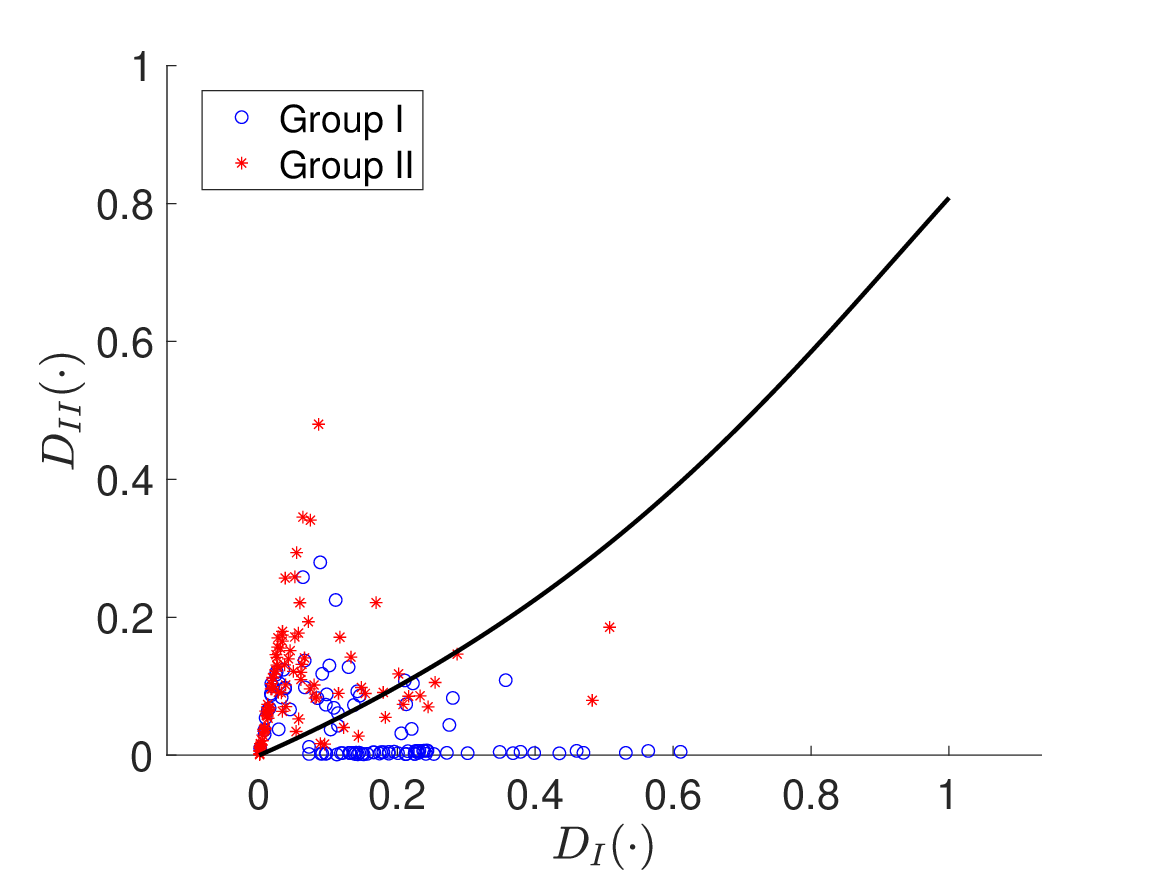}
    		\end{minipage}
    	}
	\caption{Result in real dataset 1. (a) Intensity, mean, and median in Group I. The green curve is the estimated intensity function. Each blue bar is a spike event in mean spike train, and each red bar is a spike event in median spike train.  (b) Same as (a) except for the result in Group II. (c) DD plot in the training data, where the black curve is the estimated boundary function.  (d) Same as (c) except for the test data.  }
	\label{fig:real_data1}
\end{figure*}

Analogous to the study in simulated data, we will examine the proposed methods in this real experimental dataset. To simplify the computation, we assume each spike train is an IPP realization in each group. Since the main distinction of two groups is the spike train cardinalities, we increase the hyper-parameter $r$ in the depth value in Definition \ref{def:wholedef} to put more weight on cardinality. The estimated intensity functions, as well as the mean spike trains and median spike trains, of the two groups are shown in Fig. \ref{fig:real_data1}(a) and (b), respectively.

One can find that the estimated intensity functions in the two groups differ in both magnitude and shape. The mean and median spike trains for the two groups also have apparent distinction. In each group, there is clear difference between mean and median. The classification results with the DD classifier in the training and test data are shown in Fig. \ref{fig:real_data1}(c) and (d), respectively. The ILR depth can effectively separate the two groups with the increasing boundary function. To compare our framework with other methods, the misclassification rates of the five classifiers are shown in Table. \ref{tab:real_com1}.

The result again shows the superiority of the DD classifier. For this dataset, other classifiers also show desirable performance as they all can well capture the cardinality difference between two groups.  Finally, the outliers in each training group can be detected and removed by our new method. After the outlier detection, 4 and 8 outliers (around $2\%$ of training data size) are removed from Group I and II, respectively, and the classification result in Table. \ref{tab:real_com1} shows a clear improvement in general. In conclusion, our proposed framework performs well in this real dataset, and the spike trains under different stimulation frequencies can be separated with relatively high accuracy (around 75\%).

\begin{table*}[h!]
\centering
\begin{tabular}{|c|c|c|c|c|c|}
\hline
\textbf{Classification method} & DD & MD & LM & MM1 & MM2 \\ \hline
\textbf{mis-class rate} & 0.252 & 0.272 & 0.297 & 0.287 & 0.287  \\ \hline
\textbf{mis-class rate after outlier removal} & 0.242 & 0.257 & 0.267 & 0.247 & 0.247  \\ \hline

\end{tabular}
\caption{Misclassification rates of five classifiers in real dataset 1. }
\label{tab:real_com1}
\end{table*}

\subsection{Real dataset 2: mouse ventral tegmental area}
\begin{figure*} [h!]
	\centering
	\subfigure[Group I]{
		\begin{minipage}[b]{0.47\textwidth}
			\centering
			\includegraphics[scale=0.3]{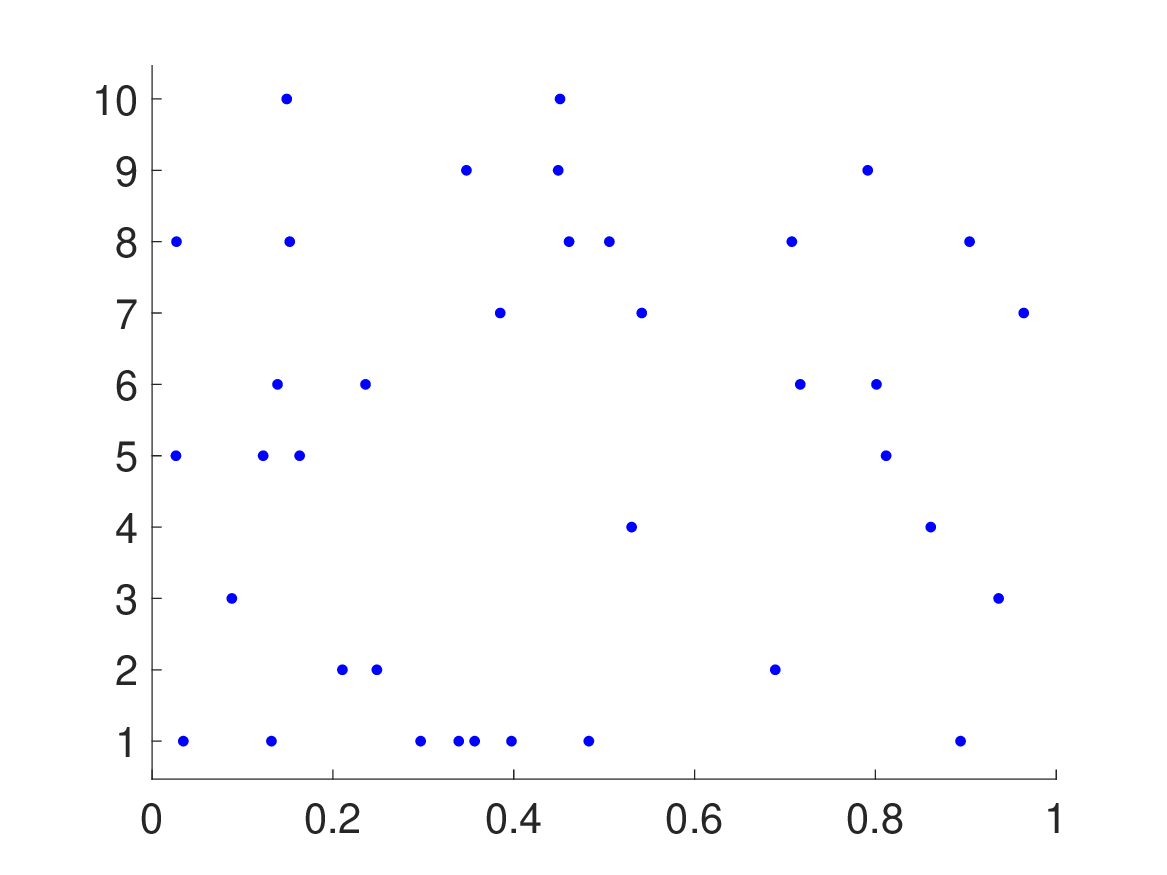}
		\end{minipage}
	}
	\subfigure[Group II]{
		\begin{minipage}[b]{0.47\textwidth}
			\centering
			\includegraphics[scale=0.3]{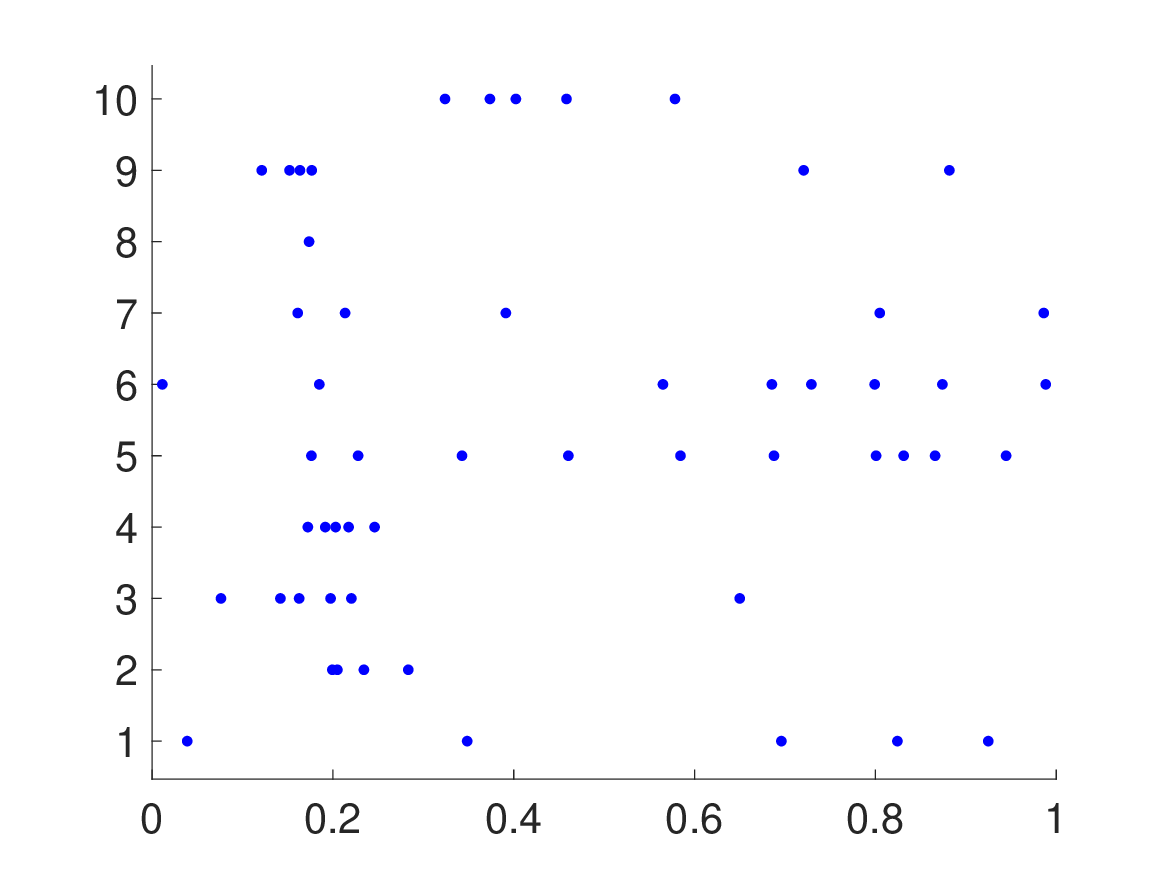}
		\end{minipage}
	}
	\caption{10 example trials in the data. (a) 10 randomly selected trials from Group I. Each row represents a trial and each blue point represents a spike event. (b) Same as (a) except for Group II. }	
	\label{fig:real_reali2}
\end{figure*}

We now demonstrate the proposed methods using another real experimental dataset. This dataset consists of spike activities of optogenetically-identified dopamine neurons in the mouse ventral tegmental area during two different conditions. In one condition, a stimulus of 1-second-long odor was provided to the animal.  In the other condition, the stimulus is water droplet deliveries for 40 ms. This experiment was first studied in \citet{starkweather2018medial} and the dataset was taken from \citet{starkweather2018spiking}. In this paper, we pick the experimental session with Salvinorin B injected to the animals. To collect two group spike trains, we first specify a fixed window length equal to the stimulus time length, and then for each trial, extract spike events immediately after the two released stimuli and within the window length. The spike trains with the same stimulus are considered as one group. In these two groups of data, the sample size is 6825 in each group. We normalize the time domain to $[0, 1]$ for computational convenience. Some example trials in the two groups are shown in Fig. \ref{fig:real_reali2}. 

\begin{figure*} [h!]
	\centering
	
	\subfigure[mean and median of Group I]{
		\begin{minipage}[b]{0.47\textwidth}
			\centering
			\includegraphics[scale=0.3]{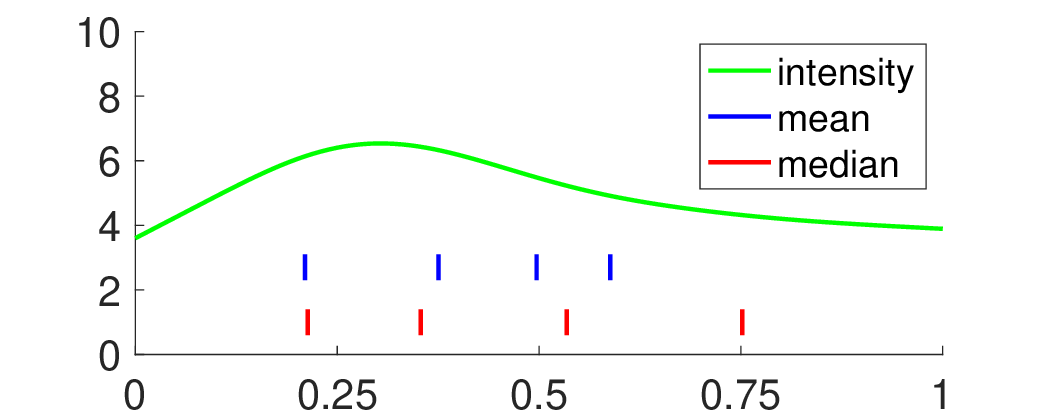}
		\end{minipage}
	}
	\subfigure[mean and median of Group II]{
		\begin{minipage}[b]{0.47\textwidth}
			\centering
			\includegraphics[scale=0.3]{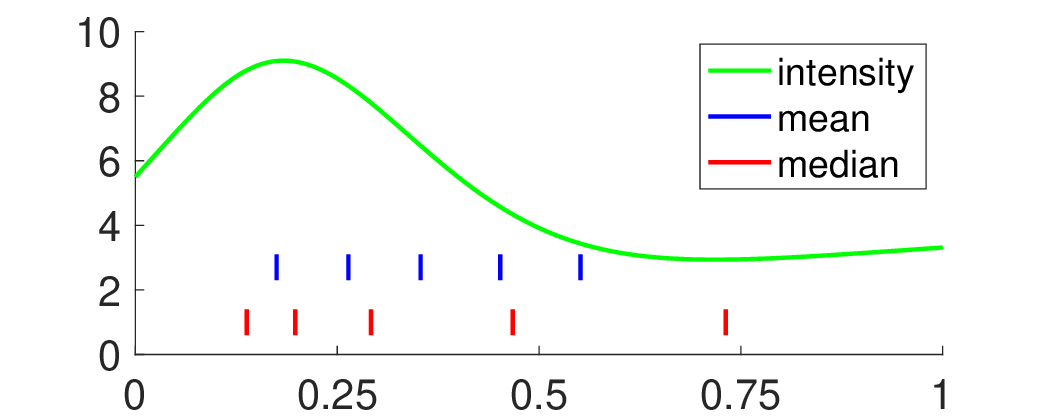}
		\end{minipage}
	}
	\\
	\subfigure[training result]{
    		\begin{minipage}[b]{0.47\textwidth}
			\centering
   		 	\includegraphics[scale=0.3]{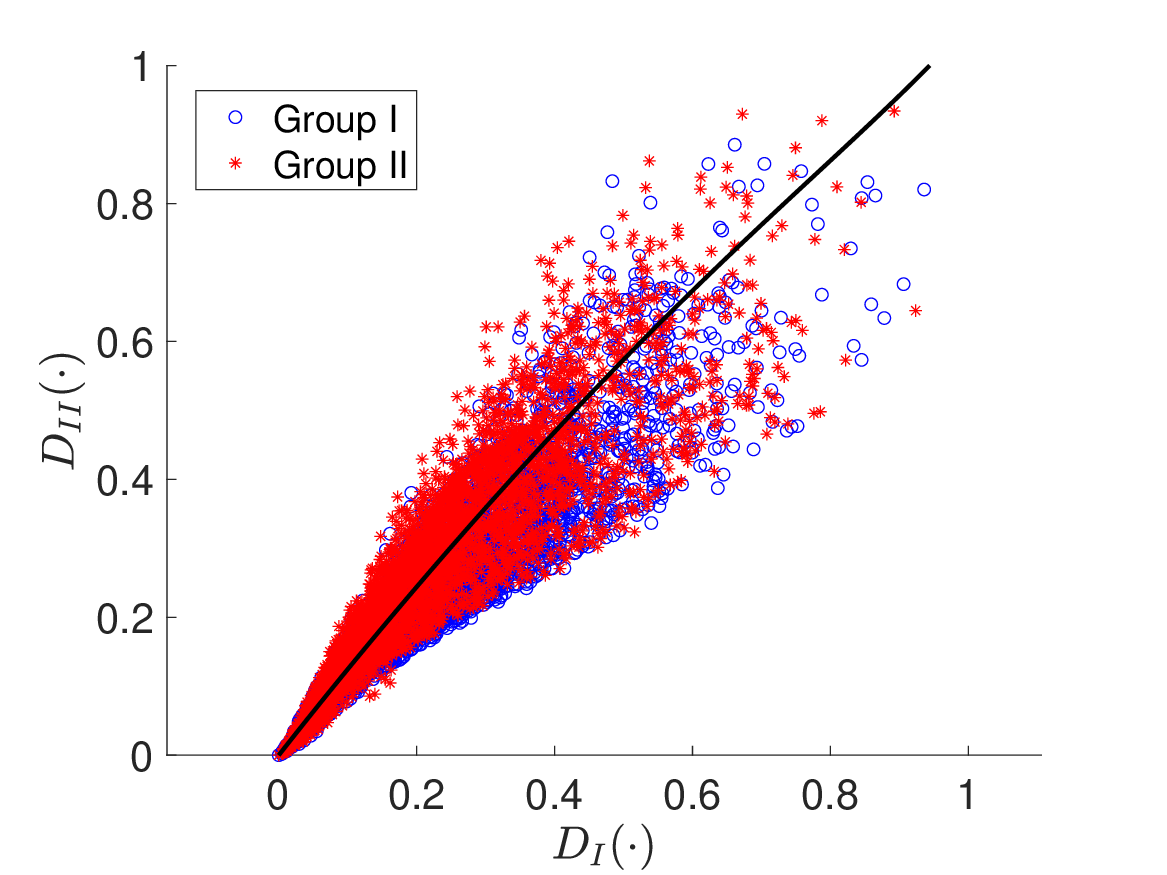}
    		\end{minipage}
    	}
	\subfigure[test result]{
    		\begin{minipage}[b]{0.47\textwidth}
			\centering
   		 	\includegraphics[scale=0.3]{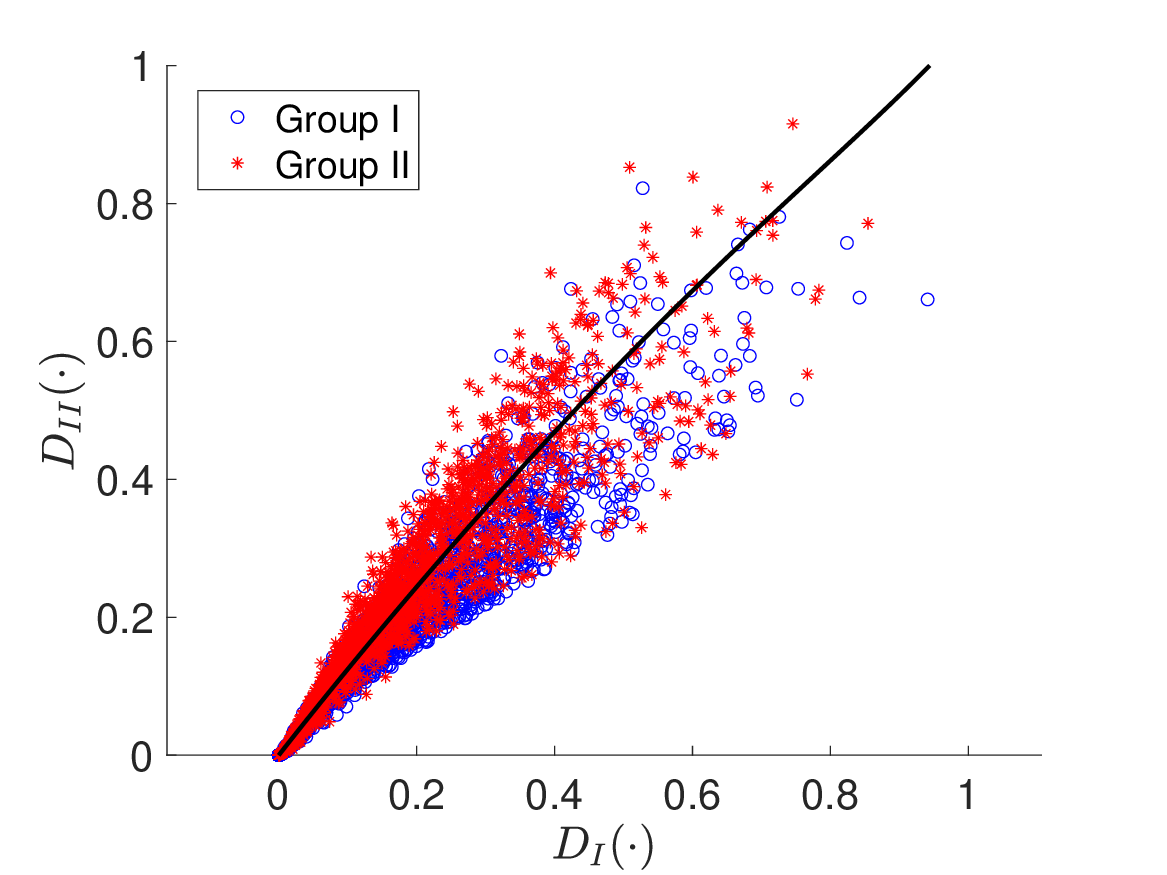}
    		\end{minipage}
    	}
	\caption{Same as in Fig. \ref{fig:real_data1} except for result in real dataset 2. }
	\label{fig:real_data2}
\end{figure*}

To implement the classification task, each group is divided into training and test set with sample sizes 5118 and 1707, respectively. In this problem, we again assume the spike trains are IPP for efficiency and robustness.  The mean and median spike trains, as well as the estimated intensity functions, for these two groups are shown in Fig. \ref{fig:real_data2}(a) and (b).   One can find that the estimated intensity function for two groups do not have obviously different patterns. The peak of the Group II is higher than that in Group I, but the two curves show similar shape. In both groups, the median spike trains show more appropriate representations for the intensities, respectively.  This indicates the median is a more reasonable template in the given data.  The DD plots on training and test data are shown in 
Fig. \ref{fig:real_data2}(c) and (d), respectively. 
We can see the DD plot does not show clear boundary to separate two groups as many points are mixed together, and this is a challenge for the proposed modified DD classifier. 

The classification result by all five methods is shown in   
Table \ref{tab:real_com2}. We can see the DD classifier still has the superior performance.  Its misclassification rate is at 0.32, lower than all other competing methods.
\begin{table*}[h!]
\centering
\begin{tabular}{|c|c|c|c|c|c|}
\hline
\textbf{Classification method} & DD & MD & LM & MM1 & MM2 \\ \hline
\textbf{mis-class rate} & 0.321 & 0.365 & 0.349 & 0.447 & 0.421  \\ \hline
\textbf{mis-class rate after outlier removal} & 0.314 & 0.346 & 0.363 & 0.437 & 0.408  \\ \hline

\end{tabular}
\caption{Misclassification rates of five classifiers in real dataset 2. }
\label{tab:real_com2}
\end{table*}
Finally, we conduct the outlier removal steps to each group.  179 and 205 are detected as outliers for Group I and II, respectively, which is around $3.5\%$ of training data size.  We then apply all classification methods to the outlier-removed data and the results are also shown in Table. \ref{tab:real_com2}. One can find the results have clear improvement in all methods except for LM.  This again supports the effectiveness of this removal procedure, which produce more accurate classification.


\section{Summary} \label{sec:summary}
In this paper, we have introduced a depth-based framework to define the notion of median for spike train sample. Similar to the property in multivariate data, the proposed median shows robustness compared with the metric-based mean spike train. In our new framework, the computation of median is also much more efficient. We have also designed a new method to detect outliers in spike train sample. This method can be conducted to any spike train model and the performance is accurate and reasonable. We have used various simulations to fully demonstrate the superiority of the proposed method. In addition, we have improved the well-known DD classifier such that its result becomes more interpretable and apply this method to conduct supervised binary classification on spike train samples. We have used both simulation examples and real experimental datasets to illustrate the feasibility of the classification method. Finally, we have systematically compared our new method with multiple competing classification methods and demonstrated the superiority of our method.

Statistical depth in spike train space has provided a principled framework to define center-outward rank in given observations.  We emphasize that the depth framework should not be limited to the proposed methods in this paper. There are other topics for further investigation in the future. For example, we can use depth-based hypothesis test to compare if two spike train samples follow the same point process model. In addition, the distribution of spike trains can be further explored with the concept of depth. One may construct a more appropriate likelihood function for spike trains to indicate the center-outward rank.  Finally, the estimation of the conditional intensity function in the ILR depth still relies on simplification (by using IPP or IMI model). More investigations are needed to capture the complexity in the intensity and better characterize the variability in the spike train data.  \\


\section*{Disclosure statement}

The authors report there are no competing interests to declare. 

\pagebreak

\noindent

\bibliographystyle{model5-names}
\bibliography{references}

\pagebreak
\appendix

\section{Derivation of threshold for outlier detection} \label{app:outlier}
Adopting Definition \ref{def:outlier} with the ILR depth, we have the following equivalence result: 
\begin{eqnarray*}
& & P\big(D(\bm{s}_0;\lambda)<t_k\big)=\delta \\
&\iff& P\big(w(k;P_{\arrowvert S\arrowvert})^rD_{\lambda}(\bm{s}_0;\arrowvert\bm{s}_0\arrowvert=k)<t_k\big)=\delta \\
&\iff& P\bigg(\frac{1}{1-\log\Big(\big(\frac{k+1}{\Lambda(T_2)}\big)^{k+1}\prod_{i=1}^{k+1}\big(\Lambda(s_i)-\Lambda(s_{i-1})\big)\Big)} <\frac{t_k}{w(k;P_{\arrowvert S\arrowvert})^r}\bigg) = \delta \\
&\iff& P\bigg(\prod_{i=1}^{k+1}\big(\Lambda(s_i)-\Lambda(s_{i-1})\big) 
<\Big(\frac{\Lambda(T_2)}{k+1}\Big)^{k+1}e^{1-\frac{w(k;P_{\arrowvert S\arrowvert})^r}{t_k}}\bigg)=\delta
\end{eqnarray*}
Since the sample of spike train is Poisson process, $\Lambda(T_2)$ is fixed for each realization. Next, based on the definition of HPP and the Time Rescaling method, $\{\Lambda(s_i)\}_{i=1}^k$ is a set of $k$ ordered uniform random variables in a fixed interval $[0,\Lambda(T_2)]$. Thus, $C_k=\Big(\frac{\Lambda(T_2)}{k+1}\Big)^{k+1}e^{1-\frac{w(k;P_{\arrowvert S\arrowvert})^r}{t_k}}$. After some simple algebra, one can easily obtain 
$$t_k=\frac{w(k;P_{\arrowvert S\arrowvert})^r}{1-\log\bigg(C_k\cdot \Big(\frac{k+1}{\Lambda(T_2)}\Big)^{k+1}\bigg)}.$$ 




\end{document}